\documentclass[12pt]{article}
\usepackage[pdfstartview=FitH,colorlinks=true,linkcolor=black,anchorcolor=black,citecolor=black,urlcolor=black]{hyperref}
\usepackage{amsthm}
\usepackage{graphicx}
\usepackage{amsmath}
\usepackage{amssymb}
\usepackage{amsfonts}
\usepackage{appendix}
\usepackage{slashed}
\usepackage{afterpage}
\usepackage{cite}
\usepackage{color}
\usepackage{pdfsync,hyperref}
\usepackage{color}
\definecolor{verde}{cmyk}{.83,.21,1,.08}
\newcommand{\rosso}[1]{{\color{black} #1}}

\newcommand{\M}{\mathcal M}

\newcommand{\sQ}{\mathsf Q}
\newcommand{\sR}{\mathsf R}
\newcommand{\sM}{\mathsf M}
\newcommand{\sN}{\mathsf N}
\newcommand{\sI}{\colorind I}
\newcommand{\sC}{\mathsf C}
\newcommand{\sD}{\mathsf D}
\newcommand{\sJ}{\colorind J}

\newcommand{\cinf}{{C^\infty({\cal M})}}
\newcommand{\complex}{{\bb C}} 

\def\alg{{\cal A}}

\font\mybb=msbm10 at 12pt
\def\bb#1{\hbox{\mybb#1}}

\newcommand{\Tr}[1]{\:{\rm Tr}\,#1}

\def\be{\begin{equation}}
\def\ee{\end{equation}}
\def\bea{\begin{eqnarray}}
\def\eea{\end{eqnarray}}

\newcommand{\del}{\partial}

\newcommand{\eqn}[1]{(\ref{#1})}

\newcommand{\spinind}[1]{\mathit{#1}}
\newcommand{\dotspinind}[1]{\mathit{\dot #1}}
\newcommand{\colorind}[1]{{\mathrm{#1}}}
\newcommand{\genind}[1]{{#1}}
\newcommand{\flavind}[1]{{\mathbf #1}}
\newcommand{\partind}[1]{{\sf #1}}

\begin{document}

\begin{titlepage}
\begin{flushright}
ICCUB-13-065
\end{flushright}

\begin{center}

\baselineskip=24pt

{\Large\bf Grand Symmetry, Spectral Action \\and the Higgs Mass}

\baselineskip=14pt

\vspace{1cm}

{Agostino Devastato$^{1,2}$, Fedele Lizzi$^{1,2,3}$ and Pierre Martinetti$^{1,2}$}
\\[6mm]
$^1${\it Dipartimento di Fisica, Universit\`{a} di Napoli
{\sl Federico II}}
\\[4mm]
$^2${\it INFN, Sezione di Napoli}\\
{\it Monte S.~Angelo, Via Cintia, 80126 Napoli, Italy}
\\[4mm]
$^3$ {\it Departament de Estructura i Constituents de la Mat\`eria,
\\Institut de Ci\'encies del Cosmos,
Universitat de Barcelona,\\
Barcelona, Catalonia, Spain}
\\{\small\tt
agostino.devastato@na.infn.it, fedele.lizzi@na.infn.it, martinetti.pierre@gmail.com}

\end{center}

\vskip 2 cm

\begin{abstract}
In the context of the spectral action and the noncommutative geometry
approach to the standard model, we  build a model based on a larger
symmetry. 
\rosso{With this \emph{grand} symmetry it
is natural to have the scalar field necessary to obtain the Higgs mass
in the vicinity of 126~GeV. This larger symmetry mixes gauge and spin degrees of freedom
without introducing extra fermions.}
 Requiring the
  noncommutative space to be an almost commutative geometry (i.e.\ the
  product of manifold by a finite dimensional internal space) gives
 conditions for the breaking of this grand symmetry to the
  standard model. 
\end{abstract}

\end{titlepage}

\section{Introduction}

Noncommutative geometry~\cite{Connesbook, Landi, Ticos,
  ConnesMarcolli}  
generalizes the concepts of ordinary geometry in an algebraic setting
and enables powerful generalizations beyond the Riemannian paradigm.   
Its application to the standard model of fundamental
interactions is a fascinating one~\cite{ConnesLott, KernerDuboisMadore, Schucker, AC2M2, Walterreview}:
the geometrical setting is that of an usual manifold
(spacetime) described by the algebra of complex valued functions
defined on it, tensor multiplied by a finite dimensional matrix algebra. This
is usually called an ``almost commutative geometry''. The standard
model is described as a particular almost commutative geometry, and
the corresponding Lagrangian is built from the spectrum of a generalized Dirac operator. 
This approach 
to the standard model has a phenomenological predictive power and is approaching the level of maturity which enables it to confront with experiments.

Schematically the application of noncommutative
geometry to the standard model has two sides. One is the mathematical
request that a topological space is a manifold. This yields a set of
algebraic requirements~\cite{Connesmanifold} involving the algebra of
functions defined on the space, represented as bounded operators on a
spinorial Hilbert space, and a (generalized) Dirac operator, plus two
more operators representing charge conjugation and chirality. These
requirements, being algebraic, can easily be generalized to noncommutative
algebra \cite{Connes:1996fu}. In the almost commutative
case \rosso{some extra assumptions on the representation
  \cite{Chamseddine:2008uq}}  single out the algebra corresponding to the standard model among a restricted number of cases~\cite{AC2M2,CCpart1} as the smallest algebra which satisfies the requirements.

The other side has to do with the spectral nature of the action.
The spectral action principle~\cite{spectralaction} allows to derive
from a unique noncommutative spacetime a Lagrangian for both
general relativity (in Euclidean signature) and the standard model. 
 The principle is purely spectral, based on the regularization
 of the eigenvalues of the Dirac operator\footnote{The spectral action
   principle, as well as any finite mode
   regularization~\cite{fujikawa,AndrianovBonora1,AndrianovBonora2},
   requires a Euclidean compact spacetime, but the cutoff on the
   momentum eigenvalues is even more general and can be used also for
   continuum spectrum, see for example~\cite{Polchinski,
     poincareinvariant}.}. 
In~\cite{AC2M2} (see also~\cite{Barrett,CCpart1}) this noncommutative model
was enhanced to include massive neutrinos and the seesaw
mechanism. The most remarkable result is the possibility to predict
the mass of the Higgs particle from the mass of the other fermions and
the value of the unification scale.
\rosso{Earlier version of the model had a prediction around 170GeV, a value ruled out by
Tevatron in 2008. Recently Connes and Chamseddine showed in~\cite{ccforgotafield} that the experimental value of the mass of Higgs at 126~GeV can be obtained
introducing a new scalar field $\sigma$ suitably coupled to
the Higgs field. Such a field had previously been proposed from a completely different
perspective by particle physicists (see for example~\cite{EliasMiro}), to avoid an high energy instability~\cite{Instability1,Instability2,Instability3} in the Higgs
potential.

The idea of a new scalar field to lower the mass of
the Higgs in the Connes approach is not new, and was already proposed by Stephan in \cite{Stephan}. However, he obtained it
adding new fermions \cite{Stephan:2005uj,Stephan:2007fa}, whereas in
\cite{ccforgotafield} (as well as in the present paper) the fermionic contents of the standard model is not
touched.  In~\cite{ccforgotafield} the field $\sigma$ is 
obtained by simply turning one of the elements of the internal Dirac operator 
into a field. 
As explained in section \ref{normalspectralaction}, this is somehow artificial because the usual
NCG procedure to obtain scalar fields (the so called
\emph{fluctuations of the metric}) does not work for the field $\sigma$,
because of one of the conditions on spectral triples (the first order
condition). 
In  \S\ref{sigma1form} we show how to overcome this difficulty by
considering a larger algebra.{\footnote{While we were preparing the second version of this paper, a proposal came out to obtain
the $\sigma$ field from a fluctuation of the metric, by relaxing the first order condition~\cite{nofirstorder,CCVPati-Salam}.}}
This is the main result of the paper.
 
More precisely, in~\cite{Chamseddine:2008uq} it is shown that under minimal
condition on the representation of the algebra,  the
smallest nontrivial almost commutative manifold corresponds to the
standard model. Here we consider a larger algebra, that we term
\emph{grand} algebra.\rosso{ We show how to obtain the field $\sigma$ by fluctuating the Majorana mass
term of the Dirac operator, in a way compatible with the first order
condition induced by this Majorana mass term. Then we show how the
first order condition imposed by the free Dirac operator reduces the grand
algebra to the one of the standard model. The field $\sigma$ then
appears as the Higgs-like field corresponding to this reduction.}
All this is possible because we intertwine in a non trivial way the
Riemann-spin degrees of freedom (the components of spinors) with the internal degrees of freedom
(the particles of the standard model). %
This puts in a new light also the phenomenon of fermion
doubling~\cite{fermiondoubling, BrunoPepeThomas, mirrorfermions,
  AC2M2} present in the theory.}

The paper is organized as follows. \rosso{In section \ref{se:spectraltriple}
we briefly recall the spectral triple cons\-truction (\S\ref{sespec}) and 
introduce the Hilbert space (\S\ref{ACmanifolds}) and the Dirac
operator (\S\ref{ACDirac}) of the standard
model. We recall in \S \ref{normalspectralaction} how the Higgs mass is obtained from the spectral
action, and point out the difficulty in generating the field $\sigma$
by fluctuation of the metric. Section
\ref{se:tripleforsm} deals with the choice of the algebras and their
representation. We first discuss the algebra of the standard model in
\S\ref{sectionasm},
then introduce the grand algebra in \S\ref{se:grandalgebra}. The reduction imposed by the grading
condition are worked out in \S\ref{gradingreduction}. 
 In section \ref{grandbreaking} we explain how the grand algebra allows
to obtain the field $\sigma$: 
first we work
out the most general Dirac operator $D_\nu$ compatible with the grand
algebra and containing a Majorana mass term for the
neutrino (\S\ref{Dirac majorana}), then we calculate
the reduction imposed by the first order condition induced by $D_\nu$ (\S\ref{majoranafirstorder}),
finally we  show that $\sigma$ can be obtained by a
fluctuation of $D_\nu$ respecting this first order
condition (\S\ref{sigma1form}). 
Section \ref{reductiontosm} deals with the reduction of the grand
algebra to the algebra of the standard model (\S
\ref{firstorderreduction}) and the issue of Lorentz invariance 
and the emergence of the spin structure (\S\ref{emergencespin}).
Possible physical interpretations are discussed in \S\ref{sectionfiat}. A final section contains conclusions and some speculative comments.}

\section{The spectral triple of the standard model} \label{se:spectraltriple}
\setcounter{equation}{0}


\subsection{Spectral triples}
\label{sespec}
The basic device in the construction is a \emph{spectral
triple} $\left(\mathcal{A},\mathcal{H},D\right)$ consisting of 
a  {*}-algebra
$\mathcal{A}$ of bounded operators in a Hilbert space $\mathcal{H}$
 - containing the identity operator -
and a non-necessarily  bounded self-adjoint operator $D$. These three
elements satisfy a set of properties allowing to prove Connes 
reconstruction theorem: 
given any spectral triple $\left(\mathcal{A},\mathcal{H},D\right)$
  with commutative $\mathcal{A}$ satisfying the required
  conditions, then ${\mathcal A}\simeq  C^{\infty}(\M)$ for some
    Riemannian spin manifold $\M$. 
\rosso{The required conditions can be found in \cite{Connesmanifold},
and their noncommutative generalization in \cite{Connes:1996fu}.} In this work we will be interested only in  
\begin{itemize}

\item[-] the \emph{grading condition}: there is an operator $\Gamma$
  (called \emph{chirality}) such
that $\Gamma^2={\mathbb I}$, $\Gamma D
= - D\Gamma$ and
\begin{equation}
[\Gamma,a]=0 \quad \forall \; a\in\alg.
\label{eq:4}
\end{equation}

\item[-] the \emph{order zero condition}: there is an antilinear isometry 
$J$ (called real structure) which implements an action of the opposite algebra{\footnote{Identical to $\alg$ as a vector space, but with reversed
product: $a^{\circ} b^{\circ} = (ba)^{\circ}$.}}
$\alg^{\circ}$ obtained by identifying
$
b^\circ = Jb^*J^{-1},
$
and which commutes with the action of $\alg$:
\begin{equation}
[a, JbJ^{-1}] = 0 \quad \forall \; a,b\in\alg.
\end{equation}
The operator $J$ must obey 
1)  $J^2 = \pm \bb I$; 2) $JD= \pm DJ$; 3)  $J\Gamma~=\pm\Gamma J$, with choice of signs dictated by the $KO$-dimension of
the spectral triple.

\item[-] the \emph{first order condition}
\begin{equation}
[[D,a], JbJ^{-1}] = 0 \quad \forall \; a,b\in\alg.
\end{equation}
\end{itemize}

\subsection{Hilbert space of the standard model \label{ACmanifolds}}

A particular form of noncommutative manifolds, suitable to describe
the standard model of elementary particles\cite{Connes:1996fu}, are the \emph{almost
  commutative} geometries, given by the product of an ordinary
  manifold $\M$ (that from now on is assumed to have dimension $4$) by
  a finite dimensional spectral triple. The
algebra is 
\begin{equation}
\label{algebra product}
\mathcal{A}=C^{\infty}(\M)\otimes\mathcal{A}_{F}
\end{equation}
where ${\cal A}_F$ is a finite dimensional algebra, whose choice is dictated by the gauge group of the standard model and
is discussed in section \ref{se:tripleforsm}.
For the Hilbert space 
a suitable choice is 
\begin{equation}
\mathcal{H}=sp(L^{2}(\M))\otimes\mathcal{H}_{F} \label{Hilbertspacetensorprod}
\end{equation}
where $sp(L^{2}(\M))$ is the Hilbert space of square summable spinors
on $\M$ and
\begin{equation}
\mathcal{H}_{F}=
\mathcal{H}_{R}\oplus\mathcal{H}_{L}\oplus\mathcal{H}_{R}^{c}\oplus\mathbb{\mathcal{H}}_{L}^{c}=\bb C^{96}
\label{eq:56}
\end{equation}
contains all the 96 particle-degrees of freedom of the standard model:
8 fermions (electron, neutrino, up and down quarks with three
colours each) for N=3 families and 2 chiralities ($\mathcal{H}_R\simeq
\mathcal{H}_L \simeq \bb C^{24}$) plus antiparticles ($\mathcal{H}_R^c\simeq
\mathcal{H}_L^c \simeq \bb C^{24}$).
The chiral and real structure are
\begin{equation}
\Gamma=\gamma^5\otimes\gamma_{F},\quad
J={\cal J}\otimes J_{F}
\label{eq:30}
\end{equation}
where ${\cal J}$ is the charge conjugation operator, $\gamma^5$
 the product of the four $\gamma$ matrices, and
\begin{equation}
\gamma_{F}=\left(\begin{array}{cccc}
\mathbb{I}_{8N}\\
 & -\mathbb{I}_{8N}\\
 &  & -\mathbb{I}_{8N}\\
 &  &  & \mathbb{I}_{8N}
\end{array}\right),\quad\,\, J_{F}=\left(\begin{array}{cc}
0 & \mathbb{I}_{16N}\\
\mathbb{I}_{16N} & 0
\end{array}\right)cc
\end{equation}
with $cc$ the complex conjugation. Notice that right
particles and left antiparticles have chirality $+1$, whereas left
particles and right antiparticles have chirality $-1$.

The Hilbert space ${\cal H}$ defined
in~\eqn{Hilbertspacetensorprod} is the tensor product of four
dimensional spinors by the 96-dimensional elements of $\mathcal H_F$,
thus (as a vector bundle over $\M$) it has dimension 384, or 128
for a single generation. This redundancy of states is known as fermion
doubling~\cite{fermiondoubling, BrunoPepeThomas, mirrorfermions}. The
problem is not only the over-counting, but the presence of states which
do not have a definite parity, being left chiral in the inner indices and right
chiral in the outer ones, or viceversa.
Since the total chirality
$\Gamma$ is the product of $\gamma_F$ (which acts on the inner indices
of $\mathcal H_F$) by $\gamma$ (which acts on the spin indices), the
spurious states are the ones for which $\Gamma\Psi=-\Psi$.
Taking the functional integral of the fermionic action to be a
Pfaffian~\cite{AC2M2} allows to project out these extra degrees of freedom. However, one cannot
simply project out the extra states and work with a representation on
a smaller Hilbert space, because in the bosonic spectral action all
degrees of freedom are necessary~\cite{fermiondoubling}
in order to obtain the proper action of the standard model coupled with gravity.
 We will see in the following that the fermion doubling may be
   in fact an essential feature of the model,  by allowing to
   represent an algebra bigger than the one of the standard model.

\subsection{Dirac operator}
\label{ACDirac}
The operator $D$ (still called Dirac operator) for the spectral
triple of the standard model~is
\begin{equation}
D=\slashed{\partial}\otimes\mathbb{I}_{96}+\gamma^5\otimes D_{F}
\label{eq:D operatore MS-1-1}
\end{equation}
with\footnote{Here $\bar{}$ denotes the complex conjugation, $\dagger$ the adjoint,
$^T$ the transpose.}
\begin{equation}
D_F=\left(\begin{array}{cccc}
0_{8N} & \mathcal{M} & \mathcal{M}_{R} & 0_{8N}\\
\mathcal{M}^{\dagger} & 0_{8N} & 0_{8N} & 0_{8N}\\
\mathcal{M}_{R} ^\dagger& 0_{8N} & 0_{8N} & \bar{\mathcal{M}}\\
0_{8N} &0_{8N} & \mathcal{M}^{T} & 0_{8N}
\end{array}\right).
\label{eq:D_F Modello Standard-1-1}
\end{equation}
The matrix $\mathcal{M}$ contains the Yukawa couplings of the fermions
and the mixing matrices (CKM for quarks and NPMS for neutrinos). It couples left
with right particles. The matrix $\mathcal{M}_{R}= {\cal M}_R^T$
contains Majorana masses and couples right particles with right
antiparticles.

The operators $\gamma_F, J_F$ and $D_F$ are such
  that
  \begin{equation}
J_F^2 =\bb I,\quad J_F D_F = D_F J_F,\quad J\gamma_F = - \gamma_F J_F,\label{eq:5}
\end{equation}
meaning that the finite part of the
  spectral triple has $KO$-dimension $6$
  \cite{AC2M2,Barrett}.  The manifold part
  has $KO$-dimension $4$, and the full spectral triple has
  $KO$-dimension $6 + 4=10 \,\,\text{mod} \,8=2$.

\subsection{Spectral action, Higgs mass and the $\sigma$ field \label{normalspectralaction}}
\rosso{Given an almost commutative geometry $(\cal A, \cal H, D)$, a \emph{fluctuation of
  the metric} \cite{Connes:1996fu} {\footnote{\rosso{The name comes from
the fact that the substitution $D\to D_A$ modifies the
      metric associated to the spectral triple. See
      \cite{Martinetti:2006db} for a detailed account on this point.}}} means the substitution of $D$ by } the gauge Dirac operator~\cite{RieffelMorita}
\begin{equation}
D_{A}\equiv D+\bb A+J\bb AJ^{-1} \label{fluctdirac}
\end{equation}
 where $\bb A=\sum_i a_i[D,b_i]$, with $a_i,b_i\in\mathcal A$,
is a generalized gauge potential. It is made
of two parts: a \rosso{scalar field on $\M$ with value in ${\cal A}_F$, and
$1$-form field on $\M$ with value in the group of unitaries of ${\cal A}_F$. In case ${\cal A}_F={\cal A}_{sm}$ is the algebra of the standard
model (discussed in \S \ref{se:tripleforsm}), the $1$-form fields yield the vector bosons mediating
the three fundamental interactions, and the scalar field is the Higgs field $H$.} 

 The spectral action \cite{spectralaction} is based on a regularization of the spectrum of $D_A$. It reads:
\be
S_B=\Tr\chi\left(\frac{D_A^2}{\Lambda^2}\right) \label{eq:spectral_action}
\ee
where $\chi$ is a cutoff function, usually the (smoothened)
characteristic function on the interval $[0,1]$, and $\Lambda$ is a
renormalization scale. It has an expansion in power series~of~$\Lambda^{-1}$, 
\be
\label{expansion}
\lim_{\Lambda\to\infty} S_B=\sum_n f_n\, a_n(D_A^2/\Lambda^2)
\ee
where the $f_n$ are the momenta of $\chi$ and  the $a_n$ are the Seeley-de Witt coefficients \cite{Gilkey,
  Vassilevich}.

Applied to the operator \eqref{eq:D operatore MS-1-1} fluctuated as in
\eqref{fluctdirac} with $a_i, b_i\in \cinf\otimes {\cal A}_{sm}$, the expansion (\ref{expansion})
yields the bosonic part of the Lagrangian of the standard model coupled with gravity~\cite[\S~4.1]{AC2M2}{\footnote{The bosonic action can also be obtained via
considerations
 related to spectral regularization and the role of
 anomalies~\cite{AndrianovLizzi,AndrianovKurkovLizzi,KurkovLizzi}.
Supersymmetric extension have been investigated in~\cite{WalterThijs}. For some cosmological predictions
based on the spectral action, see e.g.~\cite{Mairi, Mairireview,
  Matilde}.}}. Furthermore the parameters related to the Higgs come
out to be function of the parameters in $D_F$, i.e. the Yukawa
couplings, which are in turn dominated by the top quark coupling. In
this sense the model predicts the Higgs mass as a function of the
other gauge couplings, the Yukawa top mass and the scale $\Lambda$
which appears also as the scale in which the three  gauge couplings
coincide. This last point is known to be true only in an approximate sense. If one takes
the unification scale to be $\Lambda=10^{17}$GeV then one finds 
- assuming the big desert hypothesis -  a Higgs mass of the order of 170 GeV. This value is not in agreement with the recent LHC experiments~\cite{HiggsMass}.

One can think of extending the model to solve this. There have been
several proposals in this sense, and some of them are reviewed in~\cite{Schucker}. In particular C.~Stephan has proposed in~\cite{Stephan} that the presence of an extra scalar field, corresponding to the breaking of a extra U(1) symmetry, can bring down the mass of the Higgs to 126~GeV. This model however  contains extra fermions. Earlier examples of extensions are in \cite{PrisSchucker,PaschkeScheckSitarz,SchuckerZouzou,Stephan:2005uj,Stephan:2007fa,Squellari:2007zr,Stephan:2009vm}.

Recently, in \cite{ccforgotafield} the  noncommutative geometry model
was enhanced to also overcome the high energies instability of a Higgs
boson with mass around 126 GeV, in addition to predicting the correct mass. This is done ruling out the hypothesis
of the ``big desert'' and considering an additional scalar field $\sigma$
that lives at high energies and gives mass to the Majorana
neutrinos. Explicitly $\sigma$ is obtained in
  \cite{ccforgotafield} by turning (inside the finite dimensional part
  $D_F$ of the Dirac
  operator) the constant-entry $k_R$ of the
  Majorana matrix $\M_R$ into a field:
\begin{equation}
  \label{eq:68}
  k_R\to k_R\sigma
\end{equation}

However, the origin of the field $\sigma$ is quite different from the Higgs. The
latter, like the other bosons, are components of the gauge potential
$\bb A$. \rosso{They are obtained from the commutator of $D_F$ with
the algebra: $D_F$ has
constant components, that is without
space dependence, but when these numbers
are commuted with elements of the algebra they give rise to the
desired bosonic fields. One could hope to obtain $\sigma$ in a
similar way, by considering $k_R$ as a Yukawa coupling. As explained
in appendix
  \ref{appendixB}, the problem is that in taking the
commutator with elements of the algebra ${\cal A}_{sm}$, the
coefficient $k_R$ does not contribute to the
potential because of the first
  order condition.} This forced the authors
of~\cite{ccforgotafield} to ``promote to a field'' only the entry $k_R$, in a somewhat arbitrary
way. Indeed the components of $D_F$ cannot all be fields to start with, otherwise the model
would loose its predictive power, in that all Yukawa couplings would
be fields, and the masses of all fermions would run independently,
thus making any prediction impossible.
In the following (sections \ref{se:tripleforsm} and
\ref{grandbreaking}) we show that there is a way to obtain the field
$\sigma$ from $k_R$ by a fluctuation of the metric, provided one starts with
an algebra larger than the one of the standard model.

\section{Algebras and representations\label{se:tripleforsm}}
\setcounter{equation}{0} 

\rosso{Under assumptions on the representation (irreducibility, existence of
a separating vector)}, the most general finite algebra in \eqref{algebra product} that satisfies all conditions for the noncommutative space to be a manifold is
\begin{equation}
\mathcal{A}_F=\mathbb{M}_{a}(\mathbb{H})\oplus\mathbb{M}_{2a}(\mathbb{C})\quad\quad
a\in \mathbb{N}^*. \label{genericalgebra}
\end{equation}
This algebra acts on an Hilbert space of dimension $2(2a)^2$ \cite{Chamseddine:2007fk,Chamseddine:2008uq}.

\subsection{The algebra of the standard model}
\label{sectionasm}

To have a non trivial grading on $\mathbb{M}_{a}(\mathbb{H})$ the integer 
$a$ must be at least 2, meaning the simplest possibility is 
\be
{\cal A_F}= 
\mathbb{M}_{2}(\mathbb{H})\oplus\mathbb{M}_{4}(\mathbb{C}). \label{bigalgebra}
\ee
Hence an Hilbert space of dimension $2(2\cdot2)^2=32$, that is the
dimension of ${\cal H}_F$ for one generation. The grading condition $[a,\Gamma]=0$
reduces the algebra to the left-right algebra:
\be
\mathcal{A}_{LR}=\mathbb{H}_L\oplus\mathbb{H}_R\oplus\mathbb{M}_{4}(\mathbb{C}).
 \label{repa2}
\ee
This is basically a Pati-Salam model~\cite{PatiSalam}, one of the not
many models allowed by the spectral action~\cite{Constraints}. The
order one condition reduces further the algebra to~\cite{AC2M2} 
(for a review see also~\cite{Walterreview})
\be
{\cal A}_{sm}=\mathbb C\oplus \mathbb H\oplus \mathbb M_3(\mathbb C),
\ee
where $\mathbb H$ are the quaternions, which we represent as $2\times
2$ matrices, and $\mathbb M_3(\mathbb C)$ are $3\times 3$ complex
valued matrices. ${\cal A}_{sm} $ is the algebra of the standard model, that is the one whose unimodular group is
U(1)$\times$SU(2)$\times$U(3).
The details of these reductions are given in appendix \ref{appenreductionAF}.

These algebras -
tensorized by $C^\infty(\M)$ - are represented on the Hilbert space~\eqn{Hilbertspacetensorprod}, whose elements are $384$ components vectors. The number 384 comes from degrees of freedom which have different physical meaning. Some of them refer to ``internal'' degrees of freedom, like colour, some refer to the Riemannian-spin structure, and have a spacetime meaning. 
We denote a generic fermion, i.e.\ an element of $\cal H$ by
\begin{equation}
\Psi^{\mathsf C \colorind I \genind m}_{\spinind s  \dotspinind s \flavind{\alpha}} (x)\in{\mathcal H}=L^2(\M)\otimes \mathsf H_F=sp(L^{2}(\M))\otimes \mathcal H_F. \label{fullspinor}
\end{equation}
The position of the indices, whose meaning is described below, is a
matter of convention, $\Psi$ is a $\bb C^{384}$-vector valued function on
  $\M$, we write some of them as upper indices and some as lower to avoid having six indices in a row. Note the difference between $\mathsf H_F$ and $\mathcal H_F$: the
latter is a 96 dimensional space and its vectors are to be multiplied by
spinors, while the former is the larger $384$ dimensional space
which exhibits explicitly the fermion doubling over-counting. So far
in the literature the Hilbert space has been considered always in its factorized form
involving $\mathcal H_F$. One of the novelties of this work is to
use the factorized form involving $\mathsf H_F$. This allows us in section \ref{se:grandalgebra} 
to consider algebras which do not act separately on spinors and the internal part.
This means that in addition of the internal degrees of freedom
  used in \cite{CCpart1}, our tensorial notation also includes  spin
  indices $s, \dot s$.

The meaning and range of the various indices of $\Psi^{\partind C \colorind I \genind m}_{\spinind s  \dotspinind s \flavind{\alpha}} (x)$  is the following:
\begin{itemize}

\item[$\begin{array}{r}\spinind s=\spinind r,\spinind l \\
    \dotspinind s=\dotspinind 0,\dotspinind 1\end{array}$]  are
  the  spinor indices. They are not internal indices in the sense that the algebra $\cal A_F$ acts diagonally on it. They take two values each, and together they
  make the four indices on an ordinary Dirac spinor. The index
  $\spinind{s=r,l}$ indicates chirality and runs over the right, left part of the spinor, while $\dot s$ differentiates particles 
  from antiparticles. In the chiral basis one thus
  has{\footnote{The multi-index $st$ after the closing parenthesis is
      to recall that the entries of the $\gamma$'s matrices
      are labelled by indices $s,t$ taking values in the set
      $\left\{l,r\right\}$. For instance the $l$-row, $l$-column block
      of $\gamma^5$ is $\mathbb I_2$. Similarly the entries of the
      $\sigma$'s matrices are labelled by $\dot s, \dot t$ indices
      taking value in the set $\left\{\dot 0, \dot 1\right\}$: for
      instance ${\sigma^2}^{\dot 0}_{\dot 0} ={\sigma^2}^{\dot 1}_{\dot 1} = 0$. }}
\begin{equation}
  \label{eq:3}
\gamma^\mu=\left(\begin{array}{cc}  0_2
  & {\sigma^\mu}^{\dot t}_{\dot s} \\{{\overline{\sigma}}^\mu}^{\dot t}_{\dot s}&  0_2
\end{array}\right)_{st},\quad \gamma^5=\left(\begin{array}{cc}  \bb I_2
  & 0_2\\ 0_2& - \bb I_2
\end{array}\right)_{st},
\end{equation}
where for $\mu = 0,1,2,3$ one defines
\begin{equation}
  \label{eq:1}
  \sigma^\mu =\left\{\bb I _2, -i\sigma_i\right\},\quad   \bar\sigma^\mu =\left\{\bb I _2, i\sigma_i\right\}
\end{equation}
with $\sigma_{i}$, $i=1,2,3$ the Pauli matrices, namely $\sigma^0 =\bb I_2$,
\begin{equation}
\nonumber
 \label{eq:43}
\sigma^1 = - i \sigma_1 =  \left(\begin{array}{cc} 0 & -i \\ - i & 0
\end{array}\right)_{\dot s \dot t}\;
\sigma^2 = - i \sigma_2 =  \left(\begin{array}{cc} 0 & -1 \\ 1 & 0
\end{array}\right) _{\dot s \dot t}\;
\sigma^3 = - i \sigma_3 =  \left(\begin{array}{cc} -i & 0 \\ 0 & i
\end{array}\right) _{\dot s \dot t}.
\end{equation}

\item[$\colorind I =0,\ldots 3$] indicates a ``lepto-colour'' index. The zeroth ``colour'' actually identifies leptons while $\sI=1,2,3$ are the usual three colours of QCD.

\item[$\flavind \alpha=1\ldots 4$] is the flavour index. It runs over
  the set $u_R,d_R,u_L,d_L$ when $\sI=1,2,3$, and
  $\nu_R,e_R,\nu_L,e_L$ when $\sI = 0$. It repeats in the obvious way for the other generations.

\item[$\partind {C=0,1}$] indicates whether we are considering ``particles'' ($\partind{C=0}$) or ``antiparticles''  ($\partind{ C=1}$). 

\item[$\genind{m=1,2,3}$] is the generation index. The representation of the algebra of the standard model is diagonal in these indices, the Dirac operator is not, due to Cabibbo-Kobayashi-Maskawa  mixing parameters.

\end{itemize}
For the remainder of this paper the generation index $\genind m$ does
not play any role. We will therefore suppress it and work with
one generation, thus effectively considering $\mathcal H_F$ and
$\mathsf H_F$ having dimension 32 and 128 respectively.

A generic element $A=\{Q,M\}$ in $C^\infty(\M)\otimes {\cal A_F} $ (with 
$Q\in  C^\infty(\M)\otimes \bb M_2({\mathbb H})$ and $M\in
C^\infty(\M)\otimes \bb M_4(\complex)$)
acts as a matrix on vectors of $\mathsf H_F$ with index structure~\eqn{fullspinor}, it is therefore a matrix with twice as many indices\footnote{$\sD$, $\sJ$, $\beta$, have the same range as
      $\sC$, $\sI$, $\alpha$ and serve as contracting indices.}:
\be
\label{repa3}
A^{\,t \dot t \,\sC \sI\beta}_{s\dot s  \sD \sJ\alpha}= 
\delta^{\spinind{t}}_s\delta^{\dotspinind t}_{\dot s}\left(
\delta^\sC_0\delta^\sI_\sJ Q_\alpha^\beta  + \delta^\sC_1
  M^\sI_\sJ\delta_\alpha^\beta
\right) 
.\ee
Here $Q_\alpha^\beta$ evaluated at $x\in \M$ denotes the entries
  $Q_\alpha^\beta(x)\in\bb C$ of the matrix $Q(x)\in\bb M_2(\mathbb H)$, viewed as
a $4\times 4$ complex matrix with components labelled by the
  $\alpha,\beta$ flavour indices. Similarily $M^\sI_\sJ$ evaluated at
  $x$  stands for the components of the matrix $M(x)\in {\bb
    M}_4(\mathbb C)$, whose entries are labelled by the $\sI, \sJ$
  lepto-colour indices.
 
The two
  Kronecker $\delta$ at the beginning of the expression for $A$ show
  that the algebra acts in a trivial way (i.e. as the identity
    operator) on the spin indices. In other words the finite dimensional algebra
$\cal A_F$ acts only on the internal indices. The two terms in the
bracket act only on particles and antiparticles respectively, as
signified by $\delta^\sC_0$ and $\delta^\sC_1$.  They are such
that the order zero condition hold. Note in fact that for particles
the action is trivial on the $\colorind{I, J}$ indices, and for
antiparticles is trivial on the $\flavind{\alpha, \beta}$
indices. Since the real structure $J$ exchanges particles with antiparticles the two $A$ and $JBJ^{-1}$ will commute.
There is no room for the representation of a larger algebra satisfying
the order $0$ condition, unless more fermions are added, or one renounces to the trivial action on the spin indices.
The second possibility is the one we will use for the grand algebra in
the following sections.

\subsection{The grand algebra \label{se:grandalgebra}}

The case $a=3$ in \eqref{genericalgebra} 
would require a 72-dimensional Hilbert space, and there is no obvious
way to build it from the particle content of the standard model. The next case, $a=4$, requires the Hilbert space to have dimension 128, which is the dimension of $\mathsf
H_F$. Said in an other way, considering together the spin and internal degrees of freedom as part of the ``grand
Hilbert space'' $\mathsf H_F$ gives precisely the number of dimension to
represent the \emph{grand algebra}
\be
{\cal A}_G=\mathbb M_4({\mathbb H})\oplus \mathbb M_8({\mathbb C}).
\ee
\rosso{This means that $\cinf\otimes\mathcal A_G$ can be represented on the same
Hilbert space $\cal H$ as $\cinf\otimes\cal A_F$. The only difference
is that one needs to factorize $\cal H$ in \eqref{fullspinor}
as $L^2(\M)\otimes \mathsf H_F$ instead of $sp(L^2(\M))\otimes{\cal H}_F$. It is a remarkable ``coincidence''
that the passage from the standard model to the grand algebra, namely
from $a=2$ to $a'=4=2a$, requires to multiply the dimension
of the internal Hilbert space by $4$ (for $2(2a')^2= 2(4a)^2 =4(2(2a)^2)$) which
is precisely the dimension of spinors in a spacetime of dimension $4$. 
Once more we stress that no new particles are introduced: $\cal A_F$ acts on ${\cal
  H}_F={\bb C}^{32}$, $\mathcal A_G$ acts on
 $\mathsf H_F= {\bb C}^{128}$ but 
 $C^\infty(\M)\otimes {\cal A}_G$ and   $C^\infty(\M)\otimes {\cal
   A_F}$  acts on the same Hilbert space $\cal H$.}

The representation of the grand algebra $\mathcal A_G$ on $\mathsf H_F$ is more
involved than the one of $\cal A_F$ on ${\cal H}_F$ in
the previous section. 
In analogy with what was done earlier we consider an element of
${\cal A}_G$ as
two $8\times 8$ matrices, and see both of them having a block
structure of four  $4\times 4$ matrices. Thus the component $Q\in M_4(\bb H)$ of the grand algebra
gets two new extra indices with respect
  to the quaternionic component of $\cal A_F$, and the same is
  true for $M\in{\bb M}_8(\bb C)$. For the quaternions we choose to identify these two new indices with the
  spinor (anti)-particles indices $\dot 0, \dot 1$; and for the complex matrices with the spinor
  left-right indices $r,l$ introduced in \S \ref{se:tripleforsm}. This choice is not unique, and we leave a full investigation of the possible alternatives for future work. Having
  both sectors diagonal on different indices ensures
  that the order zero condition is satisfied, as explained below.

We therefore have 
  \begin{equation}
    \label{eq:16}
  Q= \left(
\begin{array}{cc}
Q_{\dot 0\alpha}^{\dot 0\beta} & Q^{\dot 1\beta}_{\dot 0\alpha}\\
Q_{\dot 1\alpha}^{\dot 0\beta} &Q_{\dot 1\alpha}^{\dot 1\beta}
\end{array}\right)_{\dot s \dot t}\in\bb M_4(\bb H) , \quad   M= \left(
\begin{array}{cc}
 M^{r\sI}_{r\sJ} &  M^{l\sI}_{r\sJ}\\
M^{r\sI}_{l\sJ} & M^{l\sI}_{l\sJ}
\end{array}\right)_{s  t}\in \bb M_8(\bb C)
  \end{equation}
where, for any
$\dot s, \dot t\in\left\{\dot 0, \dot 1\right\}$ and
$s,t\in\left\{l,r\right\}$, the matrices 
\begin{equation}
  \label{eq:82}
  Q_{\dot s\alpha}^{\dot t\beta }\in \bb M_2(\bb H),\quad
M^{t\sI}_{s \sJ} \in \bb M_4(\bb C)
\end{equation}
 have the index structure defined below \eqref{repa3}.  This means that
the representation of the element $A=(Q,M)\in{\mathcal A}_G$ is{\footnote{To
  take into account the non-diagonal action of $Q$ and $M$, it is
  convenient to change the order of the indices with respect to
  \eqn{repa3}. 
  We now adopt the order: $C,s,I,\dot s, \alpha$.}}:
\be
A^{\sC \, t \,\sI \dot t\beta}_{\sD s \sJ \dot s\alpha} = \left(
\delta^\sC_0 \delta_s^t \delta^\sI_\sJ 
\,{Q}^{\dot t \beta}_{\dot s \alpha} +
\delta^\sC_1 M^{t\sI}_{s \sJ}\delta_{\dot s}^{\dot t}\delta_\alpha^\beta\right).\label{repa4}
\ee

This representation is to be compared with 
~\eqn{repa3}.
As before the quaternionic part acts on the particle sector of the
internal indices ($\delta^\sC_0$) and the complex part
on the antiparticle sector ($\delta^\sC_1$). The difference 
is that the grand algebra acts in a nondiagonal way
not only on the flavour and lepto-colour indices $\flavind\alpha,
\colorind I$, but also on the $s$ and  $\dot s$ indices. The novelty
is in this mixing of internal and spacetime indices: at the grand algebra level, the spin structure
  is somehow hidden. \rosso{Specifically, the representation \eqref{repa4} is not
  invariant under the action of the Lorentz group (or rather of
  $Spin(4)$ since we are dealing with spin representation, in
  Euclidean signature). This point is adressed in section~\ref{emergencespin}.}

 The representation of $C^\infty(\M)\otimes{\cal A}_G$ is given
  by \eqref{repa4} where the entries of $Q$ and $M$ are now functions
  on $\M$. \rosso{Since the total Hilbert space $\cal H$ is unchanged,
    there is not reason to change the real structure and the grading. In
particular one easily checks that the order zero condition holds true for
the grand algebra
\begin{equation}
\label{orderzerogrand}
\left[A, JBJ^{-1} \right]=0\quad\forall A,B\in\mathcal{A}_{G}.
\end{equation}
This is because} the real structure $J$ in \eqref{eq:30} acts as the charge conjugation
operator 
\begin{equation}
{\cal J}= i\gamma^0\gamma^2 cc=i\left(\begin{array}{cc} {{\overline
    \sigma}^2}^{\dot t}_{\dot s}
    &0_2\\ 0_2 & {\sigma^2}^{\dot t}_{\dot s}\end{array}\right)_{s t}  \, cc
\label{eq:2}
      \end{equation}
on the spinor indices, and as $J_F$ in $\mathcal H_F$ (where it exchanges the two blocks corresponding to particles
and antiparticles). In tensorial notations one
has
\begin{equation}
(J\Psi)^{\partind C\colorind I}_{s\dotspinind s
  \flavind{\alpha}} = - i \eta^t_s\,\tau^{\dot t}_{\dot s}\, \xi^\sC_\sD \, \delta^\sI_\sJ\,
\delta^\beta_\alpha\, \bar \Psi^{\partind D \colorind J
}_{t \dotspinind t \flavind{\beta}}
\label{eq:12}
\end{equation}
where we use Einstein summation and define
\begin{equation}
  \label{eq:22}
  \xi=\left(\begin{array}{cc} 0 & 1 \\ 1 &
    0 \end{array}\right)_{\sC\sD},\quad \eta=\left(\begin{array}{cc} 1
     & 0 \\ 0&
   - 1 \end{array}\right) _{st}, \quad \tau=\left(\begin{array}{cc} 0 & -1 \\ 1 &
   0 \end{array}\right) _{\dot s \dot t}.
\end{equation}
Hence $J$ preserves the indices structure
in (\ref{repa4}), apart from the exchange
$\delta^\sC_0\leftrightarrow\delta^\sC_1$: since $Q$ and $M$ act 
on different indices, the commutation \eqref{orderzerogrand} is assured. \rosso{Notice that without the
enlargement of the action of the finite dimensional algebra to the
spinorial indices, it would have been impossible to find a
representation of $\mathcal A_G$ which satisfies the order zero condition,
unless one adds more fermions.} In this respect the grand algebra is not anymore an internal algebra.

\subsection{Reduction due to grading \label{gradingreduction} }

In a way similar to the reduction $\cal A_F\to\mathcal
A_{LR}$ explained in appendix \ref{appenreductionAF}, the grading condition imposes a reduction $\mathcal{A}_{G}\to
{\cal A}'_G$ where
 \be
{\cal A}'_G = \left(\bb M_2(\bb H)_L\oplus \bb M_2(\bb H)_R\right)\oplus \left(\bb M_4(\bb C)_l \oplus \mathbb M_4(\bb C)_r\right).
\label{AprimeG}
\ee
To see it, recall that the chirality $\Gamma$ in \eqref{eq:30} acts as $\gamma^5 = \eta_s^t \delta_{\dot s}^{\dot t}$
on the spin indices, and as $\gamma_F$  on the internal indices:
\begin{equation}
\label{tensorJgamma}
(\Gamma\Psi)^{\partind C\colorind I}_{s\dotspinind s
  \flavind{\alpha}} = \eta_s^t \delta_{\dot s}^{\dot
t}\, \,\eta^\sC_\sD\,\delta^\sI_\sJ\,\eta_\alpha^\beta \; \Psi^{\partind D\colorind J}_{t\dotspinind t
  \flavind{\beta}}
\end{equation}
where $\eta^\sC_\sD$ and $\eta^\beta_\alpha$ are defined as in
\eqref{eq:22}.
Changing the order of the indices so that to match (\ref{repa4})), one has
\begin{equation}
\label{Gammashort}
\Gamma= 
\eta^\sC_\sD \, \eta_s^t \,\delta^\sI_\sJ \,\delta_{\dot s}^{\dot t}\,\eta_\alpha^\beta.
\end{equation}
Since the representation of ${\cal A}_G$ is diagonal in the $\sC$ index, the
grading condition is satisfied if and only if it is satisfied by
both sectors - quaternionic and complex - independently.

For quaternions, one asks $[\eta_{s}^{t}\delta^{\sI}_{\sJ}\delta_{\dot
  s}^{\dot t}\eta_\alpha^\beta,
\delta_{s}^{t}\delta^\sI_\sJ Q^{\dot t\beta}_{\dot s\alpha} ]=0$, that
is $[\delta_{\dot s}^{\dot t}\eta_\alpha^\beta,
Q^{\dot t\beta}_{\dot s\alpha} ]=0$. This  imposes 
\begin{equation}
Q = 
\left(\begin{array}{cc}  
    \sQ^{\dot 0\beta}_{\dot 0\alpha} & {\sQ}^{\dot 1\beta}_{\dot 0\alpha}\\
    {\sQ}^{\dot 0\beta}_{\dot 1\alpha} & \sQ^{\dot 1\beta}_{\dot 1\alpha}
\end{array}\right)_{\dot s\dot t}
\label{eq:85}
\end{equation}
where for any $\dot s,\dot t\in\left\{\dot 0, \dot 1\right\}$ one has
\begin{equation}
\sQ^{\dot s\beta}_{\dot t\alpha}  = 
\left(\begin{array}{cc} 
    {q_R}^{\dot s}_{\dot t} & 0_2 \\ 0_2 &
    {q_L}^{\dot s}_{\dot t}
\end{array}\right)_{\alpha\beta}
\text{ with } \quad {q_R}^{\dot s}_{\dot t},    {q_L}^{\dot s}_{\dot t}\in
\bb H.
\end{equation}
Elements of the kind (\ref{eq:85}) generates $\bb M_2(\bb H)_R\oplus
\bb M_2(\bb H)_L$. Hence the reduction
\begin{equation}
\bb M_4(\bb H) \to   \bb M_2(\bb H)_R \oplus   \bb M_2(\bb H)_L.
\label{eq:54}
\end{equation}

For matrices, one asks
$[\eta_{s}^{t}\delta^{\sI}_{\sJ}\delta_{\dot s}^{\dot t}\eta_\alpha^\beta, M^{t
  \sI}_{s\sJ}\delta^{\dot t}_{\dot s}\,\delta^\beta_\alpha ]=0$, that is $[\eta_{s}^{t}\delta^\sI_\sJ , M^{t
  \sI}_{s\sJ} ]~=~0$. This forces
\begin{equation}
\label{eq:55bis}
M= \left(
\begin{array}{cc}
 M^{r\sI}_{r\sJ} &  0_4\\
0_4 & M^{l\sI}_{l\sJ}
\end{array}\right)_{s t},
 \end{equation}
meaning the reduction 
\begin{equation}
\bb M_8(\bb C) \to \bb M_4(\bb C)_{r} \oplus \bb M_4(\bb C)_{l}.
\label{eq:55}
\end{equation}
Hence the reduction of the grand algebra to ${\cal A}'_G$.
Notice that the grading causes a reduction not
 only in the quaternionic sector, as in the case of $\cal A_F$,
 but also in the complex matrix part.  This is because ${\cal A}_G$ is not anymore
 acting only on internal indices.

\section{The Majorana coupling and the $\sigma$ field \label{grandbreaking} } 
\setcounter{equation}{0} 

In this section we see how the grand algebra makes possible to
have a Majorana mass giving rise to the field $\sigma$. Although the
calculations are quite involved, the principle is quite simple. Since
we have a larger algebra, the Majorana Dirac operator needs not be diagonal in
the spin indices. This added degree of freedom enables the possibility
to satisfy the order one condition in a non trivial way, namely to
still have a one form which commutes with the opposite algebra, but
that at the same time gives rise to a field. In the following we will
show this analytically, all calculations have also been performed with
a symbolic manipulation programme, leading to the same results.

\rosso{We first work out in \S\ref{Dirac
  majorana} the most general Dirac operator $D_\nu$ with Majorana
coupling compatible with the grading condition and the $KO$ dimension
of the spectral triple of the standard model. Then we study the first order condition induced by
$D_\nu$ and the subsequent reduction $\mathcal A_G \to \mathcal A''_G$
of the grand algebra (\ref{majoranafirstorder}). Finally we show in \ref{sigma1form} that $D_\nu$ can be
fluctuated by $\mathcal A''_G$ so that to generate the field $\sigma$
as required by \eqref{eq:68}.}

\subsection{Dirac operator with Majorana mass term}
\label{Dirac majorana}
We will consider a Majorana-like mass only for the right handed
neutrinos. This choice is dictated by physics, and elsewhere we will
investigate the more general case. The natural mass scale of this
matrix is very high, so that it provides a
natural see-saw mechanism (although in realistic scheme the right handed neutrino mass is somewhat lower than the Planck scale).
The standard model can be considered as a low energy limit of the
theory we present in this section. We will assume therefore that all the
quantities involved in the internal Dirac operator $D_F$ but the Majorana coupling are small compared to the scale of the
breaking described here. \rosso{Moreover we work with one generation
  only, meaning that $D_F=D_R$ is given by \eqref{eq:60}.} We take advantage of the
flexibility introduced by the grand algebra and we do \emph{not}
assume a priori that the Majorana coupling is diagonal on the spin
indices. This means that instead of $\gamma^5\otimes D_F$ as in
\eqref{eq:D operatore MS-1-1} we consider
a finite dimensional matrix $D_\nu$ containing a Majorana mass term with non
trivial action on the spin indices.
Right handed neutrinos have indices $\colorind{I}=0$ and
$\alpha=1$, so that the most general Majorana coupling matrix is
\begin{align}
  \label{eq:9}
D_\nu = {\cal R} \otimes D_R = 
\left(\begin{array}{cc} 0_{64} & \mathsf D_\nu \\ \mathsf D_\nu^\dagger &
    0_{64}\end{array}\right)_{\sC\sD} \quad\text{ with }\quad \mathsf D_\nu = {\cal R}_{s\dot s}^{t\dot t}\;\,
\;\Xi^\sI_\sJ \;\, \Xi_\alpha^\beta 
 \end{align}
 where $\cal R$ is - at this stage - an arbitrary $4\times 4$ complex
matrix while $\Xi$ is the projector on the first component
\begin{equation}
\Xi=\left(\begin{array}{cc} 1& 0 \\ 0 &
      0_{3}\end{array}\right).
\label{eq:11}
  \end{equation}

The constraints on the matrix $\cal R$ come from the grading condition and the real
structure. Remembering
~\eqref{Gammashort}, one has that $\Xi^\sI_\sJ$ and $\Xi^\beta_\alpha$ commute with $\delta^\sI_\sJ$ and
$\eta_\alpha^\beta$, while the r.h.s. of (\ref{eq:9}) as a matrix in
$\sC\sD$ anticommutes with
$\eta^\sC_\sD$. So $D_\nu$ anticommutes with $\Gamma$ if and only if 
${\cal R}$ commutes with $\gamma^5$, meaning that
${\cal R}$ is block diagonal 
\begin{equation}
  \label{eq:10}
   {\cal R}=\left(\begin{array}{cc} {\cal R}_{r\dot s}^{r\dot t} &
       0_2\\ 0_2 &  {\cal R}_{l\dot s}^{l\dot
         t}\end{array}\right)_{st} =: \left(\begin{array}{cc} {\mathsf r}_{\dot s}^{\dot t} &
       0_2\\ 0_2 &  {\mathsf l}_{\dot s}^{\dot
         t}\end{array}\right)_{st}. 
\end{equation}
The requirement to have
$KO$-dimension $2\,\mathrm{mod}~8$ means that $J D_\nu = D_\nu J$. Remembering
(\ref{eq:12}), this is equivalent to 
\begin{equation}
  \label{eq:13}
\left[  -i\left( \begin{array}{cc} 0_4& \eta_s^t \tau_{\dot s}^{\dot t} \\ \eta_s^t
      \tau_{\dot s}^{\dot t}& 0_4
\end{array}\right)_{CD} \, cc\, ,\, \left( \begin{array}{cc}0_4 & {\cal
    R}_{s\dot s}^{t\dot t} \\  {{\cal R}^\dagger}_{s\dot
    s}^{t\dot t}& 0_4
\end{array}\right)_{CD} \right]= 0,
\end{equation}
that is
\begin{equation}
  \label{eq:14}
  (\tau \otimes\eta) {\cal R}^T - {\cal R}(\tau\otimes\eta) = 0,\quad 
  (\tau\otimes \eta) \bar{\cal R} - {\cal R}^\dagger(\tau\otimes\eta) = 0.
\end{equation}
By~(\ref{eq:10}), the first equation above yields (omitting the $st$ and $\dot s \dot t$ indices)
\begin{equation}
  \label{eq:15}
  \left(\begin{array}{cc} \tau& 0_2 \\ 0_2 &
     - \tau \end{array}\right)
\left(\begin{array}{cc} \mathsf r^T & 0_2 \\ 0_2 &
      \mathsf l^T \end{array}\right) - \left(\begin{array}{cc} \mathsf r & 0_2 \\ 0_2 &
      \mathsf l \end{array}\right)\left(\begin{array}{cc} \tau& 0_2 \\ 0_2 &
      -\tau \end{array}\right)=0 
\end{equation}
i.e.\ $\mathsf r\tau = \tau \mathsf r^T$ and $\mathsf l\tau = \tau
\mathsf l^T$, whose solution is
\begin{equation}
  \label{eq:18}
\mathsf r_{\dot s}^{\dot l} = k_r\delta^{\dotspinind l}_{\dotspinind s},\;
\mathsf l_{\dot s}^{\dot l} =k_l\delta^{\dotspinind l}_{\dotspinind s}\quad\quad k_r,
      k_l\in\bb C.
\end{equation}
The second equation in (\ref{eq:14}) is then satisfied as well.

Eq. (\ref{eq:18}), (\ref{eq:10}) and (\ref{eq:9}) give the most
general Dirac operator $D_\nu$ on $L^2(\bb R^4)\otimes \mathsf H_F$,
with Majorana mass term, coupling the right neutrino with
its anti-particle. In tensorial notations, one has
\begin{equation}
  \label{eq:21}
  \mathsf D_\nu = \kappa_s^t \, \Xi_\sJ^\sI
   \,   \delta_{\dot s}^{\dot t} \, \Xi_\alpha^\beta
\quad \text{ 
where }\quad
\kappa =\left(\begin{array}{cc} k_r & 0 \\ 0 & k_l\end{array}\right)_{st}.  
\end{equation}

By choosing $k_r = - k_l=1$, one gets ${\cal R}=\gamma^5$ and one retrieves the
  Majorana coupling $D_\nu =\gamma^5\otimes D_R$ of the standard model. However, at this stage
  nothing forces us to make this choice.

\subsection{First order condition for Majorana Dirac operator}
\label{majoranafirstorder}

We aim at obtaining the field $\sigma$ as a fluctuation of $D_\nu$, 
compatible with the first order condition. 
By (\ref{repa4})  a generic element $(Q, M)$ of ${\cal A}_G$ acts as 
{\footnote{To lighten notation, for any pairs of indices $x,y$ and $u,v$ we write $\delta^{xu}_{yv}=\delta^x_y \delta^u_v$.}}
\begin{equation}
A=\left(\begin{array}{cc}
\delta_{s\sJ}^{t\,\sI}\,  Q_{\dot s\alpha}^{\dot t \beta}& 0_{64}\\
0_{64} & M_{s\sJ}^{t\,\sI}\, \delta_{\dot s \beta}^{\dot t\alpha} 
\end{array}\right)_{\sC\sD}=:\left(\begin{array}{cc}
\mathsf Q& 0_{64}\\
0_{64} & \mathsf M
\end{array}\right)_{\sC\sD}.
\end{equation}
As well, with $B=(R, N)\in {\cal A}_G$, a
generic element of the opposite algebra is
\begin{equation}
  \label{eq:29}
  JBJ^{-1}= -J B J= - \left(\begin{array}{cc}
\tilde N^{t\sI}_{s\sJ} \, \delta_{\dot s \beta}^{\dot t\alpha} & 0_{64}\\
0_{64} & \delta_{s\sJ}^{t\,\sI}\tilde R ^{\dot t\beta}_{\dot s \beta}
\end{array}\right)_{\sC\sD}=  - \left(\begin{array}{cc}
\tilde {\mathsf N}& 0_{64}\\
0_{64} & \tilde{\mathsf R}
\end{array}\right)_{\sC\sD}
\end{equation}
where we define
\begin{equation}
  \tilde{R}_{\dot s\alpha}^{\dot t \beta} = 
  (\tau \bar R\tau)_{\dot s\alpha}^{\dot t \beta},\quad
  \tilde{N}_{s\sJ}^{t\sI} = -(\eta\bar
  N\eta)_{s\sJ}^{t\sI} = -\bar N_{s\sJ}^{t\sI}. 
\label{actionj2}
\end{equation}
  The first order condition for $D_\nu$ means that
\bea\nonumber
0= & \left[\left[D_\nu, A\right],JBJ^{-1}\right] =\left[
\left[\left(\begin{array}{cc} 0_{64} & \mathsf D_\nu \\ \mathsf D_\nu^\dagger &
    0_{64}\end{array}\right)_{\sC\sD} ,   \left(\begin{array}{cc}
\mathsf Q& 0_{64}\\
0_{64} & \mathsf M
\end{array}\right)_{\sC\sD} \right],\left(\begin{array}{cc}
\tilde{\mathsf N}& 0_{64}\\
0_{64} & \tilde{\mathsf R}
\end{array}\right)_{\sC\sD}\right]&\\
\label{eq:firordplicit}=& \left(
\begin{array}{cc} 
0_{64} & \sD_\nu \sM\tilde\sR - \sQ\sD_\nu\tilde\sR - \tilde\sN\sD_\nu\sM + \tilde\sN\sQ\sD_\nu\\ 
\sD_\nu^\dagger \sQ \tilde\sN - \sM\sD_\nu^\dagger \tilde\sN -\tilde \sR\sD_\nu^\dagger\sQ +\tilde \sR\sM\sD_\nu^\dagger& 0_{64}
\end{array}\right)_{\sC\sD}.&
\eea

We look for solutions that satisfy the grading condition, i.e. in ${\cal A'}_G$. Inspired by the first order condition for
${\cal A}_{LR}$ and $D_F$ described in appendix
\ref{appenreductionAF}, we also impose the reductions
\begin{equation}
\bb M_4(\bb C)_r \to\bb C_r\oplus\bb M_3(\bb C)_r,\quad 
\bb M_4(\bb C)_l \to\bb C_l\oplus\bb M_3(\bb C)_l
\label{eq:45}
\end{equation}
as well as
\begin{equation}
  \label{eq:71}
  \bb M_2(\bb H)_R \to \bb H_R \oplus
\bb H'_R,\quad \bb M_2(\bb H)_L \to \bb H_L \oplus
\bb H'_L.
\end{equation}
The reduction \eqref{eq:45} is obtained
assuming that the components in (\ref{eq:55bis}) are ($i,j=1,2,3$)
\begin{align}
  \label{eq:17}
  M^{r\sI}_{r\sJ} &=\left(\begin{array}{cc} M_{r0}^{r0} & 0 \\ 0&
      M_{rj}^{ri}\end{array}\right)_{\sI \sJ} =:
  \left(\begin{array}{cc} m_r & 0 \\ 0&
      M_{rj}^{ri}\end{array}\right)_{\sI \sJ} \quad m_r\in\bb
  C_r,\nonumber\\
 M^{l\sI}_{l\sJ}
  &=\left(\begin{array}{cc} M_{l0}^{l 0} & 0 \\ 0&
      M_{lj}^{li}\end{array}\right)_{\sI \sJ} 
=:\left(\begin{array}{cc} m_l & 0 \\ 0&
      M_{lj}^{li}\end{array}\right)_{\sI \sJ}\quad m_l \in
  \bb C_l.
\end{align}
The reduction \eqref{eq:71} is obtained imposing that the off-diagonal part of
$Q$ in (\ref{eq:85}) is zero:
\begin{equation}
  \label{eq:27}
 Q =
  \left(\begin{array}{cc}
\mathsf Q_{\dot 0\alpha}^{\dot 0 \beta} & 0_4 \\ 0_4 & \mathsf Q_{\dot
  1\alpha}^{\dot 1 \beta} \end{array}\right)_{\dot s\dot t}
\end{equation}
where
\begin{equation}
  \label{eq:28}
  \mathsf Q_{\dot 0\alpha}^{\dot 0 \beta} =   \left(\begin{array}{cc}
    q_R & 0_2 \\ 0_2 & q_L  \end{array}\right)_{\alpha\beta},\quad\; 
  \mathsf Q_{\dot 1\alpha}^{\dot 1 \beta} =   \left(\begin{array}{cc}
    q'_R & 0_2 \\ 0_2 & q'_L  \end{array}\right)_{\alpha\beta}
\quad\; q_{R,L}\in\bb H_{R,L},\quad q'_{R,L}\in\bb H'_{R,L}.
\end{equation}
 Finally we impose that 
$q_R$ and $q'_R$ are diagonal quaternions, that is
\begin{equation}
  \label{eq:19}
  q_R= \left(\begin{array}{cc} c_R & 0 \\ 0& \bar
      c_R\end{array}\right)_{\dot s\dot t}, \quad q'_R= \left(\begin{array}{cc} c'_R & 0 \\ 0& \bar
      c'_R\end{array}\right)_{\dot s\dot t}  \quad \text{ with } c_R, c'_R\in \bb C,
\end{equation}
meaning the reduction $\bb H_R \oplus \bb H'_R \to \bb C_R \oplus \bb C'_R$.
 We thus look for solutions of  (\ref{eq:firordplicit}) in 
 \begin{equation}
   \label{eq:24}
   {\cal A}''_G = \left(\bb H_L \oplus \bb H'_L\oplus \bb  C_R \oplus \bb C'_R \right)\oplus \left(\bb C_l \oplus \bb M_3(\bb C)_l \oplus\bb C_r\oplus\bb M_3(\bb C)_r\right).
 \end{equation}
\rosso{Notice that we do \emph{not} claim there is no solution of
  (\ref{eq:firordplicit}) outside $ {\cal A}''_G$. But for our
  purposes, it turns out that it is sufficient to work with $ {\cal
    A}''_G$.}
 
Under these conditions, the first equation coming from (\ref{eq:firordplicit}), namely
\begin{equation}
\label{eq:23}
\sD_\nu \sM\tilde\sR - \sQ\sD_\nu\tilde\sR - \tilde\sN\sD_\nu\sM + \tilde\sN\sQ\sD_\nu=0,
\end{equation}
has explicit components
{\small
\begin{eqnarray}
\sD_\nu\sM \tilde\sR =  (\kappa_s^t \, \Xi_\sJ^\sI\, \delta_{\dot s}^{\dot t} \, \Xi_\alpha^\beta)(M_{s\sJ}^{t\,\sI}\, \delta_{\dot s \beta}^{\dot t\alpha}) ( \delta_{s\sI}^{t\sJ}\tilde R_{\dot s\alpha}^{\dot t \beta})&=& (\kappa\Xi M)_{s\sJ}^{t\sI}\, 
(\Xi\tilde R)_{\dot s\alpha}^{\dot t \beta}
\nonumber\\
  &=& \left(\begin{array}{cc} k_r \,{\mathsf m_r} &  0_4\\0_4 & k_l \mathsf m_l\end{array}\right)_{s t}\otimes  
 \left(\begin{array}{cc}-\bar{\mathsf d}'_R& 0_4 \\0_4 & -\bar{\mathsf d}_R\end{array}\right)_{\dot s\dot t};
\nonumber\\
\sQ\sD_\nu\tilde\sR  =  ( \delta_{s\sI}^{t\sJ} Q_{\dot s\alpha}^{\dot t \beta}) (\kappa_s^t \, \Xi_\sJ^\sI\, \delta_{\dot s}^{\dot
    t} \, \Xi_\alpha^\beta) ( \delta_{s\sI}^{t\sJ}\tilde R_{\dot   s\alpha}^{\dot t \beta})  &= & (\kappa\Xi)_{s\sJ}^{t\sI}
 \, (Q\Xi\tilde R)_{\dot s\alpha}^{\dot t \beta} \nonumber\\
 & =& \left(\begin{array}{cc}
  k_r\;\Xi &  0_4\\
    0_4 &  k_l \;\Xi 
  \end{array}\right)_{s t}\otimes
\left(\begin{array}{cc}
-\mathsf c_R \bar{\mathsf d}'_R & 0_4 \\
     0_4 & -\mathsf c'_R \bar{\mathsf d}_R  \end{array}\right)_{\dot s\dot t}; \nonumber\\
     \tilde\sN \sD_\nu\sM 
=  ( \tilde{\mathsf N}_{s\sJ}^{t\sI}\delta_{\dot s\alpha}^{\dot t\beta}) (\kappa_s^t \, \Xi_\sJ^\sI\, \delta_{\dot s}^{\dot t} \, \Xi_\alpha^\beta) 
( \mathsf M_{s\sJ}^{t\sI}\delta_{\dot s\alpha}^{\dot t\beta}) &=&
((\tilde{\mathsf N}\kappa\Xi M)_{s\sJ}^{t\sI}\,
(\delta\Xi)_{\dot s\alpha}^{\dot t\beta} =
\nonumber\\ &=&\left(
  \begin{array}{cc} -k_r\bar{\mathsf n_r}\mathsf m_r &  0_4\\ 0_4 &- k_l\bar{\mathsf n_l}\mathsf m_l\end{array}\right)_{s t}
\otimes
\left(\begin{array}{cc} \Xi& 0 _4\\ 0_4& \Xi\end{array}\right)_{\dot s \dot t};
\nonumber\\
\tilde\sN\sQ\sD_\nu =  (\tilde\sN_{s\sJ}^{t\,\sI}\, \delta_{\dot s
  \beta}^{\dot t\alpha}) ( \delta_{s\sI}^{t\sJ} \sQ_{\dot s\alpha}^{\dot t \beta}) (\kappa_s^t \, \Xi_\sJ^\sI\, \delta_{\dot s}^{\dot t} \, \Xi_\alpha^\beta)&=  &
(\tilde \sN\kappa\Xi)_{s\sJ}^{t\,\sI}(\sQ\Xi)_{\dot s\alpha}^{\dot t \beta}=\nonumber\\
& = &\left(
  \begin{array}{cc}  -k_r\bar{\mathsf n_r}&  0_4\\  0_4 &  -k_l\bar{\mathsf
    n_l}\end{array}\right)_{s t}
\otimes  
 \left( \begin{array}{cc} \mathsf c_R & 0_4 \\ 0_4 & \mathsf c'_R\end{array}\right)_{\dot s\dot t}
\end{eqnarray}
}
where we defined the $4\times 4$ complex matrices
\begin{equation}
  \label{eq:34}
\mathsf m_{r,l} =\left(\begin{array}{cc} m_{r, l}& 0 \\ 0&0_3\end{array}\right)_{\sI \sJ}\quad
\mathsf c_{R,L} = \left(\begin{array}{cc} c_{R, L} & 0 \\
    0&0_3\end{array}\right)_{\alpha \beta}\quad
\mathsf c'_{R,L} = \left(\begin{array}{cc} c'_{R, L} & 0 \\ 0&0_3\end{array}\right)_{\alpha \beta}
\end{equation}
with $m_r, m_l$ the components of $M$ and $c_R, c'_R$ the one of
$Q$. \rosso{Similarly we define the matrices $\mathsf n_{r,l}$ from the components
$n_{l,r}$ of $N$, and the matrices $\mathsf d, \mathsf d'_R$ from the
components $d_R, d'_R$ of $\mathsf R$.  The
matrix $\Xi$ carries the indices $\sI, \sJ$ in the second equation,
and $\alpha, \beta$ in the third. In each equation, to pass from the
first to the second lines one uses \eqref{actionj2}.}

Collecting the components and assuming that both $k_r$ and $k_l$ are non zero, one
finds that (\ref{eq:23}) is
equivalent to 
\begin{align}
(c_R-m_r)(\bar n_r-\bar d'_R) = 0, &\quad  (\bar d_R - \bar n_r)(m_r- c'_R) = 0\nonumber\\
(c_R-m_l)(\bar n_l-\bar d'_R) = 0, &\quad  (\bar d_R - \bar n_l)(m_l- c'_R) = 0.
\end{align}
A similar calculation for the second components of
\eqref{eq:firordplicit} yields the same system of equations. Thus one
solution to the first order condition induced by $D_\nu$ is to impose
\begin{equation}
c_R = m_r = m_l\; \text{ and } d_R = n_r =n_l,
\label{eq:20}
\end{equation}
meaning the reduction of ${\cal A}''_G$ to 
\begin{equation}
  \label{eq:40}
  {\cal A}'''_G = \bb H_L \oplus \bb H'_L \oplus \bb C'_R \oplus \bb C \oplus \bb
  M_3(\bb C)_l \oplus \bb M_3(\bb C)_r.
\end{equation}

\subsection{The $\sigma$ field as a $1$-form}
\label{sigma1form}

We now consider the set of $1$-forms $\sum_i B_i[D_\nu, A_i]$ generated by the Majorana Dirac
operator and the algebra ${\cal A}'''_G$ above. We are interested in showing
that this set is non empty, and it is enough to consider the simplest $1$-form 
\begin{equation}
  \label{eq:42}
  [D_\nu , A]  = \left(\begin{array}{cc}
0_{64} &\mathsf D_\nu \sM - \sQ \mathsf D_\nu\\
\mathsf D_\nu^\dagger \sQ -\sM\mathsf D_\nu^\dagger& 0_{64}
\end{array}\right).
\end{equation}

We begin with $A=(Q, M)$ in ${\cal A'}_G$. With notations of the precedent section, one has
\begin{small}\begin{align*}
  \mathsf D_\nu\sM - \sQ \mathsf D _\nu &=
  (\kappa_s^t\,\Xi_\sJ^\sI\,\delta_{\dot s}^{\dot t}\,\Xi_\alpha^\beta)(M_{s\sJ}^{t\sI}\delta_{\dot s\alpha}^{\dot
    t\beta}) - (\delta_{s\sI}^{t\sJ}Q_{\dot s \alpha}^{\dot t\beta})  (\kappa_s^t\Xi_\sJ^\sI\delta_{\dot s}^{\dot
    t}\Xi_\alpha^\beta)\nonumber \\[.61em]
 &=  
    (\kappa\Xi M)_{s\sJ}^{t\sI}(\Xi\delta)_{\dot s\alpha}^{\dot
    t\beta} - (\kappa\Xi)_{s\sI}^{t\sJ}(Q\Xi)_{\dot s \alpha}^{\dot
    t\beta}\nonumber \\[.61em]
&= \left( \begin{array}{cc} k_r{\mathsf m} _r & 0_4 \\ 0_4 & k_l {\mathsf m}_ l
\end{array}\right)_{st}\otimes \left(\begin{array}{cc}
\Xi& 0_{4}\\ 0_{4}&\Xi\end{array}\right)_{\dot s\dot t} - 
\left(\begin{array}{cc} k_r \,\Xi& 0_4 \\ 0_4 & k_l\,\Xi
\end{array}\right)_{st}\otimes \left(\begin{array}{cc} \mathsf c_R & 0_4 \\
  0_4 & \mathsf c'_R 
\end{array}\right)_{\dot s\dot t}\nonumber
\\[.61em]
&= 
{\small 
\left(\begin{array}{cc}
\left(\begin{array}{cc} k_r(m_r - c_R) \Xi^{\sI\beta}_{\sJ\alpha} & 0\\ 0 &
    k_r(m_r - c'_R) \Xi^{\sI\beta}_{\sJ\alpha}\end{array}\right)_{\dot s \dot t} &
0_{32}\\
0_{32}& \left(\begin{array}{cc} k_l(m_l - 
    c_R) \Xi^{\sI\beta}_{\sJ\alpha} & 0_{32}\\ 0_{32} &
    k_l(m_l - c'_R) \Xi^{\sI\beta}_{\sJ\alpha}
 \end{array}\right)_{\dot s \dot t}
\end{array}\right)_{st}.
}
\end{align*}\end{small}
By the reduction ${\cal A'}_G\to {\cal A'''}_G$, the component
$k_r(m_r - c_R)$
vanishes, but the component $k_l(m_l - c'_R)$ does
not.  This is the crucial difference
with the algebra of the standard model: the grand algebra allows to
generates a non-vanishing $1$-form associated to the Majorana Dirac
operator $D_\nu$, which satisfies the first order condition.

Restoring the order $s\dot s\sI\alpha$ of the indices, the matrix
above is ${\cal R}= {\cal R}_{s\dot s}^{t\dot t}\,\Xi_\sI^\sJ\, \Xi_\alpha^\beta$ with
\begin{equation}
  \label{eq:25}
 {\cal R}_{s\dot s}^{t\dot t} = 
\left(\begin{array}{cc}
    \left(\begin{array}{cc}  0 &0 \\ 0& k_r (m_r-c'_R)\end{array}\right)_{\dot s\dot t} & 0_2 \\
0_2 &\left(\begin{array}{cc} 0 &0 \\ 0& k_l(m_r-c'_R)\end{array}\right)_{\dot s\dot t}  
\end{array}\right)_{st}.
 \end{equation}
 For anti-selfadjoint $A$ (that is $\mathsf M= -\mathsf M^\dagger, \mathsf Q =
 -\mathsf Q^\dagger$), one obtains the selfadjoint $1$-form
 \begin{equation}
   \label{eq:26}
    [D_\nu, A]  =\left(\begin{array}{cc}
0_{64} & \cal R \\ {\cal R}^\dagger & 0_{64}\end{array}\right).
 \end{equation}
The conjugate action of the real structure $J$ yields
\begin{equation}
  \label{eq:32}
  J   [D_\nu, A] J^{-1} = - J    [D_\nu, A] J= -\left(\begin{array}{cc}
0_{64} &  {\cal J}{\cal R}^\dagger{\cal J}\\ {\cal J}{\cal R} {\cal J}& 0_{64}\end{array}\right)
\end{equation}
where the charge conjugation ${\cal J}$ acts only on the spin
indices. Explicitly, omitting the factor $\Xi_\sI^\sJ\,\Xi_\alpha^\beta$
in the expression of $\cal R$, one gets
\begin{align}
  \label{eq:33}
   {\cal J}{\cal R}^\dagger{\cal J} &=\eta_s^t\,\tau_{\dot s}^{\dot t}\;{({\cal R}^T)}^{t\dot t}_{s\dot s} \; \eta_s^t\,\tau_{\dot s}^{\dot t}
 = \left(\begin{array}{cc}
\tau_{\dot s}^{\dot t} \,{\cal R}_{r\dot s}^{r\dot t} \,\tau_{\dot s}^{\dot t}  & 0_4 \\ 0_4 & \tau_{\dot s}^{\dot t}\, {\cal R}_{l\dot s}^{l\dot t}\, \tau_{\dot
  s}^{\dot t} \end{array}\right)_{st}\\
&=
-\left(\begin{array}{cc}
    \left(\begin{array}{cc}  k_r (m_r-c'_R) &0 \\ 0& 0\end{array}\right)_{\dot s\dot t} & 0_2 \\
0_2 &\left(\begin{array}{cc}  k_l(m_r -c'_R) &0 \\ 0&0\end{array}\right)_{\dot s\dot t}  
\end{array}\right)_{st},
\end{align}
that is  $-{\cal J}{\cal R}^\dagger{\cal J}$ is obtained by permuting
the components in the blocks $\dot s \dot t$ of ${\cal
  R}$.
As well 
\begin{equation} {\cal J}{\cal R}{\cal J} =\eta_s^t\,\tau_{\dot s}^{\dot t}\;{\bar {\cal R}}^{t\dot t}_{s\dot s} \; \eta_s^t\,\tau_{\dot s}^{\dot t}
 =\eta_s^t\,\tau_{\dot s}^{\dot t}\;({\cal R}^\dagger)^{t\dot t}_{s\dot s} \; \eta_s^t\,\tau_{\dot s}^{\dot t}
\end{equation}
is obtained from  $-{\cal R}^\dagger$ by permuting the components in
$\dot s\dot t$. Consequently,
\begin{equation}
  \label{eq:53}
  D_\nu + [D_\nu, A] + J[D_\nu, A]J^{-1}= \left(\begin{array}{cc}
      0_{64}& \cal M_\nu \\ \cal M_\nu^\dagger & 0_{64}\end{array}\right)
\end{equation}
where ${\cal M}_\nu = {\mathsf R}_{s\dot s}^{t\dot t} \Xi_{\sI}^{\sJ} \Xi_\alpha^\beta$ with
\begin{equation}
   \label{eq:58}
 {\mathsf R} = \left(\begin{array}{cc}
  k_r(1+ (m_r - c'_R) ) \delta_{\dot s}^{\dot t} & 0_2 \\
0_2 &k_l(1+(m_r - c'_R)) \delta_{\dot s}^{\dot t}
\end{array}\right)_{st}.
 \end{equation}

Now, considering that $A$ is in $C^\infty(\M)\otimes {\cal A}''_G$, the
coefficients $m_r$ and $c'_R$ becomes functions on the manifold $\M$. Taking $k_l =
-k_r= k_R$, one obtains
${\mathsf R}_{s\dot s}^{t\dot t} = k_R\sigma\gamma^5$ 
where
\begin{equation}
  \sigma = (1+(m_r-c'_R))
\label{eq:36}
  \end{equation}
is now a field on $\M$. In other terms, the fluctuation of  $D_\nu$ by
$\mathcal A_G$ \rosso{yields the substitution \eqref{eq:68}. The grand algebra gives a justification for the presence of the field $\sigma$, necessary to obtain the mass of the Higgs in agreement with experiment.}


\section{Reduction to the standard model}
\setcounter{equation}{0}
\label{reductiontosm}
\rosso{Starting with the grand algebra $\mathcal A_G$ reduced to $\mathcal
  A'_G$ by the grading condition, we have shown how to
generate the field $\sigma$ by a fluctuation of the Majorana-Dirac
operator $D_\nu$, in a way satisfying the first order condition
imposed by $D_\nu$. As explained below \eqref{eq:21}, one can choose
in particular $D_\nu=\gamma^5\otimes D_R$, where $D_R$ is the internal
Dirac operator $D_F$ of the standard model in which only the dominant term
(i.e. the Majorana mass) is taken into account. In other words, the
field $\sigma$ is generated by
fluctuating the second term in the Dirac operator \eqref{eq:D
  operatore MS-1-1} of the standard model. We now show that the first order condition of the
first term in \eqref{eq:D
  operatore MS-1-1}, i.e. the free Dirac operator, yields
the reduction of the grand algebra to the standard model.}

\subsection{First order condition for the free Dirac operator\label{firstorderreduction}}

\rosso{The
first term in \eqref{eq:D
  operatore MS-1-1} is the Euclidean free Dirac operator, extended
trivially to the internal space of one generation. In tensorial
notation it reads}  
\begin{equation}
    \label{eq:3-bis}
\slashed\del\otimes\mathbb I_{32}=-i\,\delta^{\sC\sI\beta}_{\sD\sJ\alpha}\,\gamma^\mu\del_\mu.
\end{equation}
For $A=(Q, M)\in \cinf\otimes{\cal A}'_G$, the commutator
\begin{equation}
  \label{eq:41}
  [{\slashed\del}\otimes\bb I_{32}, A] =-i\left(\begin{array}{cc}
      \delta^\sI_\sJ[ {\gamma}^\mu\del_\mu \delta_\alpha^\beta,
     \delta_s^t Q_{\dot s\alpha}^{\dot t\beta}]
      & 0_{64} \\ 0_{64} &   
[{\gamma}^\mu\del_\mu \delta^\sI_\sJ, M_{s\sJ}^{t\sI}] \delta_\alpha^\beta 
\end{array}\right)_{\partind{CD}}
\end{equation}
has components (omitting the non relevant $\delta$)
\begin{align}
  \label{eq:39}
[\gamma^\mu\del_\mu \delta^\beta_\alpha,
  \delta_s^t Q_{\dot s\alpha}^{\dot t\beta}] &=
 \left[\begin{pmatrix}  
     0_8 &{\sigma^\mu}^{\dot t}_{\dot s}\del_\mu \delta^\beta_\alpha\\ 
     {{\overline\sigma}^\mu}^{\dot t}_{\dot s}\del_\mu \delta^\beta_\alpha& 0_8
 \end{pmatrix}_{st},
 \left(
   \begin{array}{cc}
     Q_{\dot s\alpha}^{\dot t\beta} & 0_8\\
     0_8 &Q_{\dot s\alpha}^{\dot t\beta} 
   \end{array}\right)_{st}
\right]\nonumber\\
&= 
\left(
  \begin{array}{cc}
    0_8 & P_{\dot s\alpha}^{\dot t\beta}  + T_{\dot s\alpha}^{\dot t\beta,\mu}\partial_{\mu}\\
    \bar P_{\dot s\alpha}^{\dot t\beta} + \bar T_{\dot s \alpha}^{\dot t\beta,\mu}\partial_{\mu} & 0_8  \end{array}
\right)_{st}
\end{align}
where
\begin{equation}
\label{eq:39quart}
P_{\dot s\alpha}^{\dot t\beta} =  ({\sigma^\mu}^{\dot u}_{\dot s}\partial_{\mu} Q_{\dot u\alpha}^{\dot t\beta}),\quad T_{\dot s\alpha}^{\dot t\beta,\mu}  =  \left[{\sigma^\mu}^{\dot t}_{\dot s}, Q_{\dot s\alpha}^{\dot t\beta}\right]
\end{equation}
and similar definitions for $\bar P$ and $\bar T$ with $\bar\sigma$
instead of $\sigma$; and 
\begin{align}
 [\gamma^\mu\del_\mu \delta^\sI_\sJ, M^{t\sI}_{s \sJ}\delta_{\dot s}^{\dot t}] &= 
\left[\left(
   \begin{array}{cc}  
     0_{8} &{\sigma^\mu}ì^{\dot t}_{\dot s}\del_\mu \delta^\sI_\sJ\\ 
     {{\overline\sigma}^\mu}^{\dot t}_{\dot s}\del_\mu \delta^\sI_\sJ& 0_{8}
   \end{array}
\right)_{st},\left(
  \begin{array}{cc}
    M^{r\sI}_{r\sJ} \delta_{\dot s}^{\dot t}&0_{8}\\
    0_{8} &M^{l\sI}_{l\sJ} \delta_{\dot s}^{\dot t}
  \end{array}\right)_{st}\right]\nonumber\\
 \label{eq:39ter}
&=
\left(
  \begin{array}{cc}
    0_{8} & L^{\sI\dot t}_{\sJ\dot s} + K^{\sI\dot t,\mu}_{\sJ\dot s} \partial_\mu \\
    \bar L^{\sI \dot t}_{\sJ\dot s} +  \bar K^{\sI\dot t,\mu}_{\sJ\dot s}\partial_\mu  &  0_{8}
  \end{array}\right)_{st}
\end{align}
where
\begin{align}
  L^{\sI\dot t}_{\sJ\dot s} &= \left({\sigma^\mu}^{\dot t}_{\dot s}
    \del_\mu M^{l\sI}_{l\sJ}\right) ,\quad  
  K^{\sI\dot t,\mu}_{\sJ\dot s} =\left(M^{l\sI}_{l\sJ}-
    M^{r\sI}_{r\sJ}\right){\sigma^\mu}^{\dot t}_{\dot s},\nonumber\\
   \bar L^{\sI\dot t}_{\sJ\dot s} &= \left({{\overline\sigma}^\mu}^{\dot t}_{\dot s}\del_\mu M^{r\sI}_{r\sJ}\right) ,\quad  \bar K^{\sI\dot t,\mu}_{\sJ\dot s} =
    \left(M^{r\sI}_{r\sJ}-
      M^{l\sI}_{l\sJ}\right){{\overline\sigma}^\mu}^{\dot t}_{\dot s}.
\end{align}

For $B=(R,N)\in{\cal A}'_G$, the commutator of $[\slashed\del, A]$
with $JBJ$ given in \eqref{eq:29} is
a block diagonal matrix in $\partind{CD}$ with components
{\small 
\begin{align}
\label{eq:36new}
&&\left[\delta^\sI_\sJ\,[ {\gamma}^\mu\del_\mu \delta_\alpha^\beta,
  \delta_s^t Q_{\dot s\alpha}^{\dot t\beta}], {\tilde N}^{t\sI}_{s\sJ}\,\delta_{\dot s\alpha}^{\dot t\beta}\right]=
\nonumber\\
&&=\left[\left(
  \begin{array}{cc}
    0_{32} & \delta^\sI_\sJ (P_{\dot s\alpha}^{\dot t\beta}  +  T_{\dot s\alpha}^{\dot t\beta,\mu}\partial_{\mu})\\
     \delta^\sI_\sJ(\bar P^{\dot t \beta}_{\dot s \alpha}  + \bar
     T_{\dot s \alpha}^{\dot t\beta, \mu}\partial_{\mu} )& 0_{32}
  \end{array}
\right)_{st}, \left(
  \begin{array}{cc}
    \tilde N^{r\sI}_{r\sJ}\delta_{\dot s \alpha}^{\dot t \beta}&0_{32} \\ 
    0_{32}&  \tilde N^{l\sI}_{l\sJ}\delta_{\dot s \alpha}^{\dot t\beta}
  \end{array}\right)_{st}\right];\nonumber\\
&&~\nonumber\\
&&\left[[{\gamma}^\mu\del_\mu \delta_\sJ^\sI,
     M_{s\sJ}^{t\sI}\delta_{\dot s}^{\dot t}]\,\delta_\alpha^\beta,\;\delta^{t\sI}_{s\sJ}\,{\tilde R}^{\dot t\beta}_{\dot s\alpha}
   \right]=\nonumber\\
   &&= \left[
\left(
  \begin{array}{cc}
    0_{32} & (L^{\sI\dot t}_{\sJ\dot s} + K^{\sI\dot t,\mu}_{\sJ\dot s} \partial_\mu )\delta_\alpha^\beta\\
    (\bar L^{\sI \dot t}_{\sJ\dot s} +  \bar K^{\sI\dot t,\mu}_{\sJ\dot s}\partial_\mu)  \delta_\alpha^\beta&  0_{32}
  \end{array}\right)_{st},
\left(
  \begin{array}{cc} \delta_\sI^\sJ \,\tilde R^{\dot t\beta}_{\dot s\alpha}& 0_{32} \\ 
0_{32}&\delta_\sI^\sJ \,\tilde R^{\dot t\beta}_{\dot s\alpha}
  \end{array}\right)_{st}
\right].
\end{align}
}
Omitting the indices (and noticing that the $P, T, \bar P, \bar T$ all
commute with $\tilde N^r_r, \tilde N^l_l$), the first components is a
diagonal matrix with first entry
\begin{equation}
  \label{eq:89}
  (\tilde N^l_l - \tilde N^r_r)(P + T^\mu\del_\mu) + T^\mu(\partial_\mu
  \tilde N^l_l).
\end{equation}
The vanishing of the differential operator part implies either $T^\mu
=0$ or $\tilde N^l_l = \tilde N^r_r$. But the expression should be zero in particular for non-constant fields, that is
for $P\neq 0$. So in case one imposes $T^\mu=0$, the vanishing of the term in $P$
implies $\tilde N^l = \tilde N^r$. In case one imposes $\tilde N^l_l =
\tilde N^r_r$, the vanishing of the remaining term implies either
$T^\mu=0$, or $\tilde N^l_l =\tilde N^r_r = \text{cst}$. The last solution
is unacceptable, it would mean that spacetime is reduced to a point, hence in any case one has both conditions:
$T^\mu = 0$ and $\tilde N^l_l =
\tilde N^r_r$. One then checks that the other components of
(\ref{eq:36new}) vanish as well.

 The only matrix that commutes with all the Pauli matrices
is the identity, therefore
\begin{equation}
  \label{eq:90}
  T^\mu=0 \; \forall \mu \Longleftrightarrow  Q^{\dot 0\beta}_{\dot
    0\alpha} = Q^{\dot 1\beta}_{\dot 1\alpha} \text{ and } Q^{\dot
    0\beta}_{\dot 1 \alpha} = Q^{\dot 1\beta}_{\dot 0\alpha}=0, 
\end{equation}
meaning the breaking 
\begin{equation}
\bb M_2(\bb H)_L \oplus \bb M_2(\bb H)_R \to \bb H_L \oplus \bb H_R.\label{eq:91}
\end{equation}
Meanwhile $\tilde N^l_l =
\tilde N^r_r$ means that
\begin{equation}
\bb M_4(\bb C)_l \oplus \bb M_4(\bb C)_r \to \mathbb M_4(\bb C).
\label{eq:92}
\end{equation}
Thus
\begin{equation}
{\cal A}'_G\to\bb H_L
\oplus \bb H_R \oplus \bb M_4(\bb C)
\label{eq:59}
\end{equation}
where representation of the r.h.s. algebra is now diagonal on the spinorial
indices $\dot s, s$.
\newline

To summarize, the grand algebra ${\cal A}_G$ is broken by the
chirality and the first order condition of the free Dirac operator to the left-right symmetric
algebra  ${\cal A}_{LR}$ of the standard model.

\subsection{Emergence of spin \label{emergencespin}}

In noncommutative geometry the topological  information is encoded in
the algebra, while the geometry (e.g.\ the metric\footnote{The
  metric aspects of the almost commutative geometry of the standard
  model have been investigated in\cite{spectralaction,PierreRaimar}}) is in the $D$ operator.
In particular the Riemann-spin structure is encoded in the way this
operator, which contains the gamma matrices, acts on the Hilbert
space. Without this operator there is
just an algebra which acts in an highly reducible way on a 128
dimensional Hilbert space. This is conceptually what distinguishes $\mathcal H_F$ from
$\mathsf H_F$ in~\eqn{fullspinor}: on  $C^\infty(\M)\otimes \mathsf {\cal H}
_F$, the free Dirac operator (trivially extended to the internal
indices)   is
\begin{equation}
  \label{eq:8}
  \slashed \del = -i\gamma^\mu\partial_\mu \otimes \delta^{\sC\sI\beta}_{\sD\sJ\alpha}.
\end{equation}
On   $C^\infty(\M)\otimes \mathsf H_F$
the same operator writes
  \begin{equation}
    \label{eq:7}
    \slashed \del = -i\partial_\mu \otimes \delta^{\sC\sI\beta}_{\sD\sJ\alpha}\gamma^\mu
  \end{equation}
and the spin structure, carried by the $\gamma$ matrices, is hidden among the
internal degrees of freedom. In this sense the first order condition, which  governs the passage
from \eqref{eq:7} to \eqref{eq:8}, corresponds to the emergence of the
spin structure.

Alternatively, a spin structure means that the vectors in the
Hilbert space transform in a particular representation under the
``Lorentz'' group. Since we are dealing with spinors in the Euclidean
case, the group is actually 
Spin(4). \rosso{It is generated by the commutators of the Dirac
  matrices, that act on $\cal H$ as 
  \begin{equation}
S^{\mu\nu} := [\gamma^\mu, \gamma^\nu]\otimes \mathbb I_{32}^{(\sC\sI\alpha)}.\label{eq:69}
\end{equation}
Let us distinguish between an element $a$ of $\cinf\otimes\mathcal
A_G$ and
its representation $\pi(a):=A$ given in \eqref{repa4}. For any
$\Lambda=\lambda_{\mu\nu}S^{\mu\nu}\in \text{Spin}(4)$
and $A\in\pi(\cinf\otimes\mathcal A_G)$, let
\begin{equation}
\alpha_\Lambda A = U(\Lambda) A U(\Lambda)^*.\label{eq:70}
\end{equation}
The representation~\eqn{repa4} of the grand algebra is not
invariant under the adjoint action \eqref{eq:70} of Spin(4) since $\alpha_\Lambda
\pi(a)$ is not in $\pi(\cinf\otimes\mathcal A_G)$. In this sense the representation of the grand algebra is not
Lorentz invariant, unlike its reduction to ${\cal A}_{LR}$ which is
diagonal in the spin indices. However, at the abstract level the algebra is preserved
under Lorentz transformations since the latter are implemented by unitary
operators: for any $\Lambda$ one has that  
$\alpha_\Lambda(\pi(\cinf\otimes\mathcal A_G))$ is isomorphic to $\cinf\otimes\mathcal
A_G$. This suggests to view 
the grand algebra as a phase of the universe in which
the spin and rotation structure of spacetime has not yet
emerged, only the topology (i.e.\ the abstract algebra) is fixed.}


\subsection{Fiat neutrino}
\label{sectionfiat}
The grand algebra together with the Majorana Dirac operator $D_\nu$ generates
the field $\sigma$ at the right position (\rosso{i.e.\ as required in \eqref{eq:68}}), respecting the first order
condition induced by $D_\nu$. However, by \eqref{eq:36} one has that $\sigma$ becomes constant when one takes into account the
first order condition imposed by the free Dirac operator, because
\eqref{eq:90} implies that $c'_R = c_R= m_r$. This suggests a scenario
in which the neutrino Majorana mass is the first field to appear \rosso{and
fluctuate, before the geometric structure of
spacetime emerges through the breaking described in \S
\ref{emergencespin}. In this picture, the field $\sigma$ is viewed as a
fluctuation of a vacuum that satisfies the first order condition of
the free Dirac operator.

This scenario is supported by some preliminary calculations, which indicate} that the first order
condition of the free Dirac operator can be equivalently obtained as a
minimum of the spectral action. In this way, the \rosso{geometrical}
breaking imposed by the mathematical requirement of the theory becomes
a dynamical breaking, and the field $\sigma$ would appear as the ``Higgs
field'' associated to it. This idea has been investigated,
in the case of the standard model algebra, in the recent papers
\cite{nofirstorder, CCVPati-Salam}. The case of the grand algebra is
in progress.

\section{Conclusions and outlook}
\label{conclusion}
\setcounter{equation}{0}

It is known that, although the spectral action requires the
unification of interactions at a single scale, the usual grand unified
theories, such as SU(5) or SO(10), do not fit in the noncommutative
geometry framework, and are possible only renouncing to
associativity~\cite{Raimarnonassociative, FansworthBoyle}. In this
paper  we pointed out there is a ``next level'' in noncommutative geometry, but that it is intertwined with the Riemannian and spin structure of spacetime, and therefore it naturally appears at a high scale.
The added degrees of freedom are related to the Riemann-spin structure
of spacetime, which emerges as a symmetry breaking very similar in
nature to the Higgs mechanism. In addition, the higher symmetry 
explains the presence of the $\sigma$ field necessary for a correct
fit of the mass of the Higgs. The results presented
here, as is common in this model, are crucially depending on the
Euclidean structure of the theory. This is particularly important as
far as the role of chirality and the doubling of the degrees of freedom is concerned. A Wick rotation is far from simple in this context, and the construction of a Minkowskian noncommutative geometry is yet to come (for recent works see~\cite{Verch, Franco}).

The presence of this grand symmetry will have also phenomenological
consequences which should be investigated. The breaking mechanisms
described in this paper  are just barely sketched, we only looked at the group structure. A more punctual analysis should reveal more structure, and possibly alter the running of the constants at high energy.

For the moment we can only speculate. One of the problems of the spectral action in its present form is that it requires unification of the three gauge couplings at a single scale, $\Lambda$, and physical predictions are based on the value of this scale. It is known experimentally that in the absence of new physics the three constants do not meet in a single point, but the  three lines form an elongated triangle spanning nearly three orders of magnitude. On the other side in the spectral action is not clear what would happen after this point, if one consider scales higher that $\Lambda$, i.e. earlier epochs. For a theory dealing with the unification of gauge theory and gravity a more natural scale is the Planck scale. An unification of the coupling constants at the Planck scale in the form of a pole has been considered~\cite{MaianiParisiPetronzio, ULP}, but it requires new fermions. In the case at hand the ``new physics'' is in the form of a different structure which mixes spacetime spin and gauge degrees of freedom. This might have consequences for the interactions, and hence for the running of the various quantities, as well.

Other mathematical issues should be investigated.  In particular the
choice of the action of the grand algebra on the spin indices reflects
how much the algebra is not diagonal in the ``interaction'' encoded by
the free Dirac operator. Hopefully this could be interpreted at the
light of Connes unitary invariant in Riemannian geometry\cite{connesckm}. 

\smallskip

\paragraph{Acknowledgments}
We would like to thank M.A.~Kurkov and D.~Vassilevich for discussions. 
F.~Lizzi acknowledges
support by CUR Generalitat de Catalunya under project
FPA2010-20807. P.M. warmly thanks A. Roche for constant support.

\appendix
\section{Reduction of the $\cal A_F$ algebra}
\label{appenreductionAF}

We give the details of the reduction of $\cal A_F$ to ${\cal
  A}_{sm}$ by the grading and the first order condition. Rather than ${\cal H}_R \oplus {\cal H}_L \oplus {\cal H}_R^c
\oplus {\cal H}_L^c$, it is convenient to work in the $\sC \sI\alpha$ basis
of ${\cal H}_F= {\bb C}^{32}$ (one generation of leptons $l$ and quarks $q$), namely
  \begin{equation}
  \underset{\sC=0}{\underbrace{
    \underset{\sI=0; \,\alpha=1, ...,4}{\underbrace{
          {\cal H}_{lR}\oplus {\cal H}_{lL}}
      }
    \oplus
    \underset{\sI=i;\, \alpha=1, ...,4}{\underbrace{
         ({\cal H}_{qR}\oplus {\cal H}_{qL})\otimes \bb I_3}
       }}
     } 
\oplus  \underset{\sC=1}{\underbrace{
    \underset{\sI=0; \,\alpha=1, ..., 4}{\underbrace{
          {\cal H}^c_{lR}\oplus {\cal H}^c_{lL}}
      }
    \oplus
    \underset{\sI=i;\, \alpha=1, ..., 4}{\underbrace{
         ({\cal H}^c_{qR}\oplus {\cal H}^c_{qL})\otimes \bb I_3.}
       }}
     } 
\label{eq:62}
  \end{equation}
In this basis the internal Dirac operator is 
\begin{equation}
  \label{eq:47}
  D_F=\left( \begin{array}{cc} D^1_1 & D^1_2\\ D^2_1 = {D^1_2}^\dagger
      &D^2_2 =  \bar{D^1_1}\end{array}\right)_{\sC\sD}
\end{equation}
where
\begin{equation}
  \label{eq:48}
  D^1_1= \left( \begin{array}{cc} {\cal M}^0 & 0 \\ 0 & {\cal
        M}^i\end{array}\right)_{\sI\sJ}, \quad  D^1_2= \left( \begin{array}{cc} {\cal M}^R & 0 \\ 0 & 0_{12}\end{array}\right)_{\sI\sJ}
\end{equation}
are $16\times 16$ matrix with 
\begin{equation}
  \label{eq:37bis}
{\cal M}^0=  \left(\begin{array}{cccc}
0 & 0 & \bar k_{\nu} & 0\\
0 & 0 & 0 & \bar k_{e}\\
k_{\nu} & 0 & 0 & 0\\
0 & k_{e} & 0 & 0
\end{array}\right)_{\alpha\beta}, \quad {\cal M}^{i=1,2,3}= \left(\begin{array}{cccc}
0 & 0 & \bar k_{u} & 0\\
0 & 0 & 0 & \bar k_{d}\\
k_{u} & 0 & 0 & 0\\
0 & k_{d} & 0 & 0
\end{array}\right) _{\alpha\beta}, \quad {\cal
M}^R= \left(\begin{array}{cccc}
k_R & 0 & 0 & 0\\
0 & 0 & 0 & 0\\
0& 0 & 0 & 0\\
0 & 0 & 0 & 0
\end{array}\right) _{\alpha\beta}
\end{equation}
\rosso{where $k_e, k_u,k_d, k_\nu$ are the Yukawa couplings of the electrons,
quarks and neutrino, and $k_R$ is the neutrino Majorana mass.}

Let us first examine the grading condition.  By (\ref{repa3}) the element $A=(Q, M)\in{\cal A_F}$ 
act on ${\cal H}_F$ as
\begin{equation}
A=\left(\begin{array}{cc}
\mathsf Q& 0_{16}\\
0_{16} & \mathsf M 
\end{array}\right)_{\sC\sD}
\end{equation}
where
\begin{equation}
\mathsf Q=\delta^\sI_\sJ Q_\alpha^\beta\in{\bb M}_2(\bb H),
\quad \mathsf M = M^\sI_\sJ\delta_\alpha^\beta\in{\bb M}_4(\bb C).
\end{equation}
To guarantee that $A$ commutes with $\Gamma$, the factor $\eta_\alpha^\beta$ in \eqref{tensorJgamma} 
requires $Q$ to be diagonal in the $\alpha$ index, which  reduces this part of the algebra to $\mathbb H_L\oplus \mathbb
 H_R$. This means that $Q^\alpha_\beta$ in~\eqn{repa3}  acts separately on
 the left and right handed doublets.
 No such breaking occurs in complex part, because $\delta^\sI_\sJ$ in \eqref{tensorJgamma} does not
 put any constraints on $\bb M_4(\bb C)$. Likewise, $\eta_s^t$ does not produce any breaking
 because the action of both quaternions and complex matrices is diagonal on the spin indices. Thus we have ${\cal A_F}\to{\mathcal A}_{LR}$. 

Let us now examine the first order condition. For any $B=(R,N)\in{\cal A}_{LR}$ one has
\begin{equation} 
J_FBJ_F=\left(\begin{array}{cc}
\bar{\mathsf N} & 0_{16}\\
0_{16} & \bar{\mathsf R}
\end{array}\right)_{\sC\sD}.
\end{equation}
Assuming there is no neutrino Majorana mass (i.e $D^1_2=0)$, the first order condition for the finite dimensional spectral
triple yields
\begin{equation}
 \left[\left[D_F, A\right], J_FBJ_F\right]=
\left(\begin{array}{cc}
 \left[\left[D^1_1, \mathsf Q\right], \bar{\mathsf N}\right] & 0_{16}\\
 0_{16} & \left[\left[\bar{D^1_1}, \mathsf M\right], \bar{\mathsf R}\right]
\end{array}\right)_{\sC\sD}=0.
\label{eq:first order explicit}
\end{equation}
Let $n^\sI_\sJ$ be the components of $\bar{N}\in\bb M_4(\bb C)$. The upper-left term in the r.h.s. of \eqref{eq:first order explicit}
is
\begin{equation}
  \label{eq:38}
\left[\left(\begin{array}{cc}
[{\cal M}^0, Q]_\alpha^\beta & 0\\
0 & [{\cal M}^i, Q]_\alpha^\beta\otimes {\bb I_3}
\end{array}\right)_{IJ}, \left(\begin{array}{cc}
  n^0_0\delta_\alpha^\beta & n^0_j\delta_\alpha^\beta \\
  n^i_0\delta_\alpha^\beta &
  n^i_0\delta_\alpha^\beta \end{array}\right)_{IJ} \right].
\end{equation}
It is zero if and only if
\begin{equation}
  \label{eq:46}
  n^0_i [{\cal M}^0 -{\cal M}^i,Q]_\alpha^\beta  =  n_0^i [{\cal M}^0 -{\cal M}^i,Q]_\alpha^\beta  = 0 \quad \forall i=1,2,3. 
\end{equation}
Writing
$Q=\left(\begin{array}{cc} q_R&0 \\0& q_L\end{array}\right)\in
\bb H_L \oplus \bb H_R$
with
$q_R=\left(\begin{array}{cc} a_1 & a_2\\ -\bar a_2 &\bar
    a_1\end{array}\right)$ and $\quad q_L=\left(\begin{array}{cc} a_3 & a_4\\ -\bar a_4 &\bar
    a_3\end{array}\right)$
 this means
\begin{align}
n^0_i(a_1-a_3)(k_\nu-  k_u) = n^0_i (a_1-a_3)(k_e-  k_d) & =0\,,\,\, i=1,2,3\nonumber\\
n^0_i \left(a_2(k_\nu - k_u) - a_4(k_e - k_d) \right)=  n^0_i\left(
a_2(k_e - k_d) - a_4(k_\nu - k_u)\right)& =0\,,\,\, i=1,2,3
\end{align}
and a similar equation for $n^i_0$. 

A first solution could be $a_1=a_3$ and $a_2=a_4=0$, which means that
  the quaternionic part reduces to $\bb C$ while $\bb M_4(\bb C)$ is not
  touched. The gauge group is then $U(1)\times U(4)$, which is to
  small to contain the gauge group of the standard model. The other
  solution is imposing $n^i_0=n^0_i=0$, which yields the reduction
  $\mathbb{M}_{4}(\mathbb{C})\rightarrow\mathbb{C}\oplus\mathbb{M}_{3}(\mathbb{C})$. Then
  the second component of (\ref{eq:first order explicit}) vanishes as
  well. Thus the first order condition without 
Majorana mass, together with the grading condition, induces the breaking:
\begin{equation}
  \label{eq:84bis}
  {\cal A_F} \to \mathcal A_{LR}\to \left(\bb H_L \oplus \bb H_R\right) \oplus \left(\bb C \oplus \bb M_3(\bb C)\right). 
\end{equation}

A non-zero Majorana coupling $k_{R}$ (a constant at this stage) in the Dirac operator
induces new terms in (\ref{eq:first order explicit}): 
\begin{eqnarray} 
  \label{eq:49}
&\left[\left[  \left(\begin{array}{cc} 
0_{16}  &D^1_2 \\ {D^1_2}^\dagger & 0_{16} 
\end{array}\right), \left(\begin{array}{cc} 
\mathsf Q  & 0_{16}  \\ 0_{16}&\mathsf M  
\end{array}\right)\right], \left(\begin{array}{cc} 
\bar{\mathsf N}  &0_{16} \\ 0_{16} &\bar{\mathsf R} 
\end{array}\right)\right] \\
\nonumber
&=\left(\begin{array}{cc} 0_{16} & D^1_2\sM\bar\sR - \sQ D^1_2\bar\sR - \bar\sN D^1_2\sM
    + \bar\sN \sQ D^1_2\\
{D^1_2}^\dagger \sQ \bar\sN - \sM {D^1_2}^\dagger \bar\sN - \bar\sR {D^1_2}^\dagger
\sQ + \bar\sR\sM {D^1_2}^\dagger&0_{16} 
\end{array}\right).
\end{eqnarray}
\rosso{Writing
$\bar\sR=\left(\begin{array}{cc} q'_R&0 \\0& q'_L\end{array}\right)\in
\bb H_L \oplus \bb H_R$
with $q'_R=\left(\begin{array}{cc} b_1 & b_2\\ -\bar b_2 &\bar
    b_1\end{array}\right)$, $q'_L=\left(\begin{array}{cc} b_3 & b_4\\ -\bar b_4 &\bar
    b_3\end{array}\right)$,}
one gets
\begin{small}
\begin{align}
  \label{eq:50}
  D^1_2\sM\bar\sR &= \left(
    \begin{array}{cc} 
      ({\cal M}^R \bar R)_\alpha^\beta &0\\0&0_{12}
    \end{array}\right)_{IJ}
  \left(\begin{array}{cc}
      m^0_0\delta_\alpha^\beta &0\\
      0&m^i_j\delta_\alpha^\beta
    \end{array}\right)_{IJ}\\
  &= \left(\begin{array}{cc} 
      m^0_0({\cal M}^R \bar R_\alpha^\beta)&0\\0&0_{12}
    \end{array}\right)_{IJ}
  = \left(
    \begin{array}{cc} 
      k_R\left(\begin{array}{cc} 
         \left(\begin{array}{cc}
           m^0_0 b_1 & m^0_0 b_2\\ 0&0
         \end{array}\right) 
         & 0_2 \\ 
         0_2 & 0_2 
       \end{array}\right)_{\alpha\beta} 
     &0\\0&0_{12}
   \end{array}\right)_{IJ},
\end{align}
and similarly 
\begin{align}
\sQ D^1_2\bar\sR &= \left(\begin{array}{cc} 
 k_R\left(
\begin{array}{cc}
  \left(\begin{array}{cc} a_1 b_1 & a_1 b_2\\ -\bar a_2 b_1&
      -\bar a_2 b_2
    \end{array}\right)  & 0_2\\ 0_2&0_2 \end{array}\right)_{\alpha\beta}
  &0\\0&0_{12}
\end{array}\right)_{IJ},
\end{align}
\begin{align}
\bar\sN D^1_2\sM &= \left(\begin{array}{cc} 
 k_R
\left(\begin{array}{cc}
    \left(\begin{array}{cc} m^0_0 n^0_0 & 0\\ 0&
        0\end{array}\right) &0_2\\ 0_2&0_2\end{array}\right)_{\alpha\beta} 
     &0\\0&0_{12}
\end{array}\right)_{IJ},
\end{align}
\begin{align}
\label{eq:50bis}
\bar\sN \sQ D^1_2&= \left(\begin{array}{cc} 
 k_R\left(\begin{array}{cc} 
\left(\begin{array}{cc}
    n_0^0 a_1 & 0\\ -n_0^0\bar a_2 &
    0\end{array}\right) & 0_2 \\ 0_2 & 0_2\end{array}\right)_{\alpha\beta}
     &0\\0&0_{12}
\end{array}\right)_{IJ}.
\end{align}
\end{small}
Asking (\ref{eq:50}) to be zero is thus equivalent to the system
\begin{align}
m^0_0 b_1 - a_1b_1 - m^0_0n^0_0 + n^0_0 a_1 =(m^0_0-a_1)( b_1- n^0_0)
= 0  \label{firstorun}\\
b_2(m^0_0 - a_1 )=0,\quad 
\bar a_2 b_1  - n^0_0\bar a_2 = 0, \quad
\bar a_2 b_2 = 0,
\label{firstordeux}
\end{align}
leading to $a_{1}=m^0_0$, $b_{1}=n^0_0$ and $a_{2}= b_2 = 0$.  This means
 \begin{equation} 
\left(\mathbb{H}_L\oplus\mathbb{H}_R\right)\oplus\left(\mathbb{C}\oplus\mathbb{M}_{3}(\mathbb{C})\right)\rightarrow
\left(\mathbb{H}_L \oplus \mathbb{C}'\right)\oplus\left(\mathbb{C}\oplus\mathbb{M}_{3}(\mathbb{C})\right)
 \label{eq:51}
 \end{equation}
with
\begin{equation}
\mathbb{C}=\mathbb{C}'.\label{eq:35}
\end{equation}
Hence the standard model algebra
$\mathcal{A}_{sm}=\mathbb{C}\oplus\mathbb{H}\oplus\mathbb{M}_{3}(\mathbb{C}).$
\section{Fluctuation of $D_R$ by the standard model algebra}
\label{appendixB}
\setcounter{equation}{0}

The notations are the one of appendix A. One easily checks that the set of $1$-forms induced by the Majorana Dirac operator 
\begin{equation}
  \label{eq:60}
  D_R =\left(\begin{array}{cc} 0_{16} & D_2^1\\ {D_2^1}^\dagger & 0_{16}\end{array}\right)_{\sC\sD} 
\end{equation}
is actually zero. \rosso{Indeed, an element of ${\cal A}_{sm}$ is
\begin{equation}
A=(\sQ, \sM)\quad \text{ with }\quad  \sQ=\delta^\sI_\sJ
Q_\alpha^\beta, \quad \sM=M^\sI_\sJ\delta_\alpha^\beta
\end{equation}
 where $Q_\alpha^\beta$ is as below \eqref{eq:46} with $a_2=0$, and
 $M$ has components $m^i_0=
m^0_i=0, i=1,2,3.$}
One thus gets
\begin{equation}
  \label{eq:65}
  [D_R , A] = \left(
\begin{array}{cc}
0_{16}& D^1_2 \sM - \sQ D^1_2\\ {D^1_2}^\dag \sQ - \sM {D^1_2}^\dag &
0_{16}
\end{array}
\right)_{\sC\sD}
\end{equation}
with
\begin{equation*}
  \label{eq:66}
   D^1_2 \sM - \sQ D^1_2 = 
\left(\begin{array}{cc}
    ({{\cal M}_R}m_1^1 )_\alpha^\beta - (Q{\cal M}_R)^\beta_\alpha  & 0\\ 
    0& 0_{12}
\end{array}\right)_{\sI\sJ} = 
\left(\begin{array}{cc}
    \left(\begin{array}{cc} 
    k_R (m^1_1-a_1) & 0 \\ 0&0_3 
  \end{array}\right)_{\alpha\beta} & 0\\
0 & 0_{12}
\end{array}\right)_{\sI\sJ}
\end{equation*}
which vanishes because of \eqref{eq:35}. The same is true for ${D^1_2}^\dagger\sQ - \sM {D_1^2}^\dagger$. 
Hence
\begin{equation}
[D_R,A]=0.\label{eq:6}
\end{equation}

\rosso{One may think of inverting the order of the reductions: first
  impose the first order condition of the Majorana-Dirac operator
  $D_R$, then the one of $D_0:=D_F-D_R$. By repeating the computation (\ref{eq:50})-~(\ref{eq:50bis}) with
$A,B$ elements of ${\cal A}_{LR}$ (instead of being elements of the
algebra at the r.h.s. of \eqref{eq:84bis}),} one obtains extra-constraints 
  \begin{align}
    \label{eq:64}
    b_1 m_i^0 - m^0_0 n_i^0 &= 0\nonumber\\
-m_0^i n_0^0 + n_0^i a_1 &= 0\nonumber\\
m_0^i n_i^0 = 0
  \end{align}
whose solution is $m^i_0 = m^0_i=0$.  This means that
the breaking
\begin{equation}
  \label{eq:63}
  {\cal A}_{LR}\to {\cal A}_{sm}
\end{equation}
can also be obtained directly from $D_R$
alone, without reducing first to the algebra
\eqref{eq:84bis}.

Consequently, starting from $C^\infty(M)\otimes{\cal A_F}$ reduced to $C^\infty(M)\otimes{\cal A}_{LR}$ by the grading condition,
there is no way to fluctuate the Dirac operator - respecting the
first order condition - so that to obtain the field $\sigma$ as
required by eq. \eqref{eq:68}.


\end{document}